\documentclass[final,1p,times]{elsarticle}

\makeatletter
\def\ps@pprintTitle{%
    \let\@oddhead\@empty
    \let\@evenhead\@empty
    \let\@oddfoot\@empty
    \let\@evenfoot\@empty
}
\makeatother

\usepackage{amssymb}
\usepackage{amsmath}
\usepackage{hyperref}
\hypersetup{
  colorlinks=true,
  allcolors=blue}
\newcommand{\bluecite}[1]{\textcolor{blue}{\cite{#1}}}
\usepackage{graphicx}
\usepackage{subcaption}
\usepackage{array}
\usepackage{pdflscape}
\usepackage{comment}

\begin{document}

\begin{frontmatter}



\title{Internet of Harvester Nano Things: A Future Prospects}


\author[label1]{Bitop Maitra\corref{cor1}}
\ead{bmaitra23@ku.edu.tr}
\author[label1]{Emine Bardakci}
\ead{ebardakci20@ku.edu.tr}
\author[label2]{Oktay Cetinkaya}
\ead{oktay.cetinkaya@ncl.ac.uk}
\author[label1,label3]{Ozgur B. Akan}
\ead{akan@ku.edu.tr, oba21@cam.ac.uk}
\cortext[cor1]{Corresponding author}

\affiliation[label1]{organization={Center for neXt-generation Communications (CXC), Department of Electrical and Electronics Engineering, Koç University},
            city={Istanbul},
            postcode={34450}, 
            country={Turkey}}

\affiliation[label2]{organization={The School of Engineering, Newcastle University},
            city={Newcastle upon Tyne},
            postcode={NE1 7RU}, 
            country={United Kingdom}}

\affiliation[label3]{organization={Internet of Everything (IoE) Group, Electrical Engineering Division, Department of Engineering, University of Cambridge},
            city={Cambridge},
            postcode={CB3 0FA}, 
            country={UK}}

\begin{abstract}
The advancements in nanotechnology, material science, and electrical engineering have shrunk the sizes of electronic devices down to the micro/nanoscale. 
This brings the opportunity of developing the Internet of Nano Things (IoNT), an extension of the Internet of Things (IoT). 
With nanodevices, numerous new possibilities emerge in the biomedical, military fields, and industrial products. 
However, a continuous energy supply is mandatory for these devices to work. 
At the micro/nanoscale, batteries cannot supply this demand due to size limitations and the limited energy contained in the batteries. 
Internet of Harvester Nano Things (IoHNT), a concept of Energy Harvesting (EH) integrated with wireless power transmission (WPT) techniques, converts the existing different energy sources into electrical energy and transmits to IoNT nodes. 
As IoHNTs are not directly attached to IoNTs, it gives flexibility in size. 
However, we define the size of IoHNTs as up to 10 cm.
In this review, we comprehensively investigate the available energy sources and EH principles to wirelessly power IoNTs.
We discuss the IoHNT principles, material selections, and state-of-the-art applications of each energy source for different sectoral applications.
The different technologies of WPT and how communication is influenced by the incorporation of IoHNTs to power IoNTs are discussed with the future research directions.
IoHNTs represent a shift in the nanodevice power supply, leading us towards a future where wireless technology is widespread. 
Hence, it will motivate researchers to envision and contribute to advancing the following power revolution in IoNT, providing unmatched simplicity and efficiency.
\end{abstract}




\end{frontmatter}



\section{Introduction} 
\label{sec:Intro}
The Internet of Things (IoT), which has made significant progress in recent years, enables smart devices to exchange information autonomously over the Internet. 
Based on the recent advances in nanotechnology and communication engineering at the micro/nanoscale, another paradigm called the Internet of Nano Things (IoNT) emerges as a novel research topic. 
Since the IoNT devices are only a few nanometers in size, they can be used in almost every field, such as healthcare, military, security, sensing, processing, actuating, and communication. 
In theory, for healthcare applications, IoNTs can be implanted inside the human body to collect health related parameters and then transmit them to healthcare providers via the Internet. 
Healthcare providers decide which commands to execute, such as synthesizing and releasing drugs \text{{\color{blue} \cite{akyildiz2015internet}}}. 
However, realizing this implementation is challenging due to the health-related risks that the nanodevices can cause to living organisms, and biocompatibility is questionable. 
Hence, the practical usage of IoNT in humans is still in its infancy. 
In that sense, the IoNT has evolved into the Internet of Bio-Nano Things (IoBNT), alleviating the biocompatibility problem and using nanodevices practically. 
However, micro/nanodevices developed accordingly have scarce processing, memory, and network capabilities, which prevent them from performing complex tasks individually. 
Instead of making a nanodevice perform a complex task by itself, dividing the task into subparts and assigning each part to a nanodevice can be done. 
The task can be realized if energy can be transmitted continuously to IoNTs, and communication between the nanodevices can be ensured.
Therefore, it is necessary to widely deploy these micro/nanodevices in a network area called ``nanonetwork".
\renewcommand{\figurename}{Fig.}

\begin{figure*}[t]
\centering
{\includegraphics[width=\textwidth]{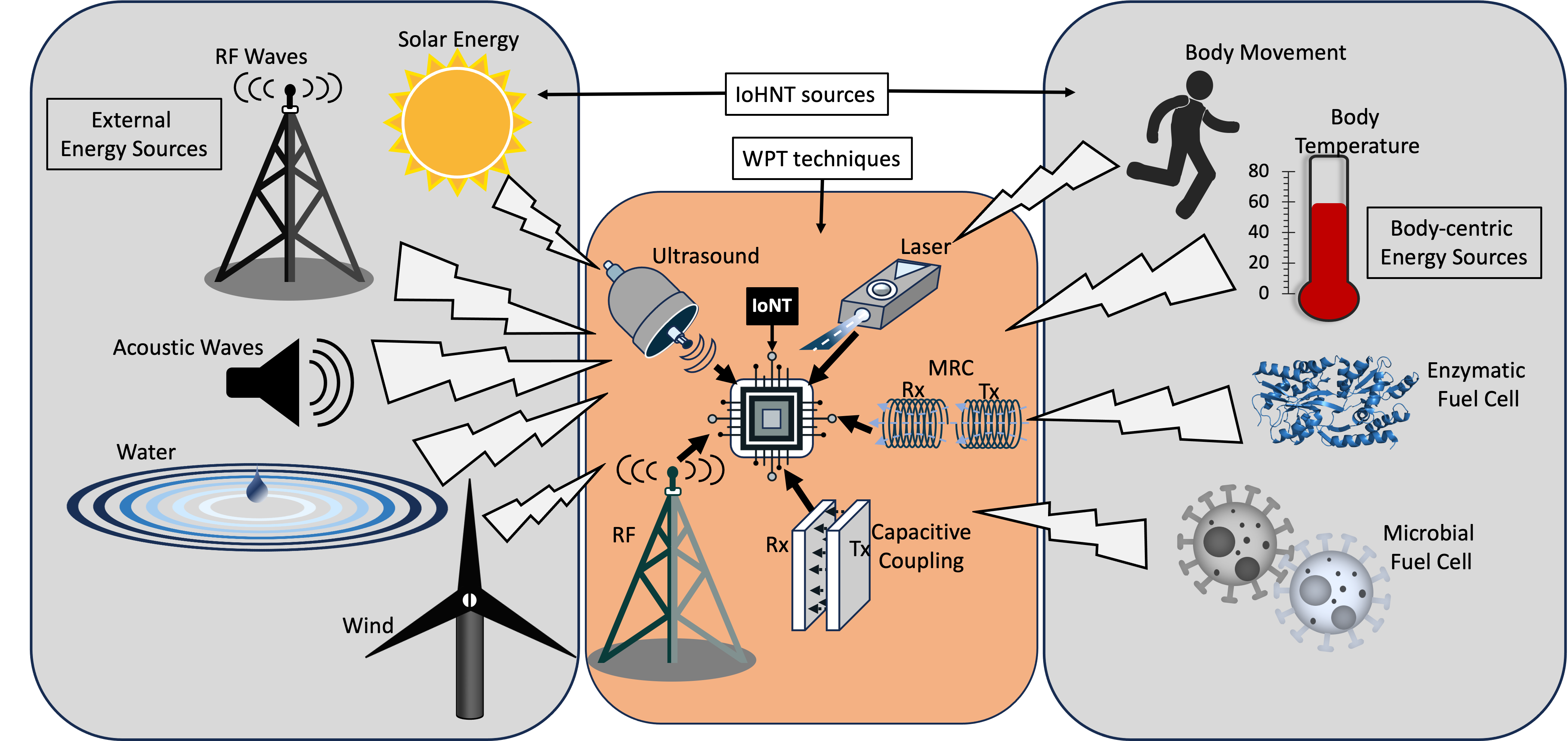}}
\caption{An overview of the Internet of Harvester Nano Things (IoHNTs).}
\label{summary}
\end{figure*}

The increase interest in these fields causes the concept of IoHNT to become more prominent since the micro/nanodevices constructed by this paradigm work more suitably than bulky batteries. 
In addition, their usage is not proper due to the biocompatibility issue \text{{\color{blue} \cite{armand2008building}}}. 
IoHNTs can convert energy sources into electrical energy in the EH system. 
The harvested energy can be directly transmitted to IoNTs through wireless power transfer (WPT) techniques, making the size of IoHNTs flexible.
Hence, IoHNT is comprised of two technologies: one is EH, and the other one is WPT. 
The power consumption of nanodevices is typically in $\sim$$\mu$W range \bluecite{Nanotechnology}, sometimes even lower to nW range \bluecite{Klauk2007-sk}.
Here, we define the maximum size of IoHNTs as up to 10 centimeters, as it can produce a significant amount of power for the continuous and advanced operation of IoNTs.  

IoNTs are micro/nanoscale devices comprised of nanosensors and actuators, communication protocols, data processing and storage, network infrastructure, a control unit, and a communication unit \bluecite{alabdulatif2023internet}, usually administered by a conventional battery. 
IoNTs, which are 1 nm to a few hundred of $\mu$m in size, cannot accommodate a significant space for a large power battery, leads to a reduction in operation time, gives rise to frequent battery replacement.
Hence, the battery must be replaced frequently after a certain period to ensure continuous operation. 
The tiny size of IoNTs and their frequent placement in inaccessible locations make the replacement operations challenging in many cases \bluecite{SHAIKH20161041}.
Therefore, there is an urgent requirement for EH mechanisms that can scavenge energy and power IoNTs, leading to a push towards an environment-friendly and cost-effective method for IoNTs \bluecite{altinel2019modeling}.
Even though there is some proposed research to power IoT devices \bluecite{garg2017energy, kamalinejad2015wireless}, there is no significant progress to power the IoNTs.
Thus, in this review, we aim to construct a readable synthesis of IoHNTs by providing external, body-centric, and hybrid EH, and how those IoHNTs can be used to power the IoNTs wirelessly by reviewing the existing EHs and proposing the extension to IoNT domain. A conceptual overview is depicted in Fig. \ref{summary}.

This article classifies the energy sources according to where they originate. 
If the energy appears in or on the human body, it is named \textit{body-centric energy source}, otherwise named as \textit{external energy source}. 
The external energy sources are categorized into flow and radiant-based energy sources. 
Flow-based energy sources are the flow of water and air, whereas radiant-based energy sources are waves (light, radio, and acoustic). 
A pictorial illustration of IoHNTs is depicted in Fig. \ref{summary}.
For classification purposes, we have considered the body-centric as the human body; however, the produced IoHNT's applicability can be extended to universal purposes.
Furthermore, we categorize body-centric energy sources, whether the human body provides the energy directly or originates using specific mechanisms, e.g., biofuel cells. 

A Wireless Power Transfer (WPT) system is the most efficient technology for transmitting energy to an IoNT. 
The energy that IoHNT produces is required to send efficiently to the functioning module. 
Until now, a battery-based nanodevice has been used frequently, but an efficient WPT system needs to be replaced for miniaturization. 
For in-vivo applications, it will be a ground-breaking approach to use efficient WPT, drastically reducing surgical hazards and improving the longevity of IoNTs.

In summary, we have explored and reviewed all the existing EH techniques. Adaption of the IoHNT technique depends upon the energy availability and based on the application. 
After discussing all the available EH techniques, we reviewed the use of wireless power transfer (WPT) to transmit the harvested power to IoNTs efficiently. 
Hence, integration of WPT in IoHNTs leads to a network where IoHNTs can efficiently communicate and transmit energy to nearby IoNTs or IoHNTs. 
These types of networks will have a significant impact in the future. 
This review ends with discussing how all the IoNT nodes can be powered continuously to be functional with better communication architecture by incorporating higher data rates, enhanced security, better hopping techniques, etc.

The rest of this paper is organized as follows. 
The external and body-centric EH sources and techniques are introduced in Sec. \ref{sec:External Energy Sources and Harvesting} and in Sec. \ref{sec:Body-centric Energy Sources and Harvesting}, respectively. 
Then, hybrid EH techniques are discussed in Sec. \ref{sec:Hybrid Energy Harvesting}. 
Next, wireless power transfer (WPT) techniques that can be combined with IoHNTs are explained in Sec. \ref{sec:Discussion}. 
Later, the impact of IoHNTs in IoNT's communication is explored in Sec. \ref{sec:influ} along with its future scope in Sec. \ref{sec:future}.
Finally, Sec. \ref{sec:Conclusion} contains a comprehensive conclusion with the required improvement in making efficient IoHNT to seamlessly power IoNT.

\section{External Energy Sources and Harvesting}
\label{sec:External Energy Sources and Harvesting}
The environment itself contains different energy sources. 
However, a considerable amount of energy the Sun delivers to the Earth goes to waste. 
The external energy sources can be divided into radiant (light, radio waves, and acoustic waves) and flow-based (wind and water). 
As external energy is abundant, generating electrical energy for nanonetworks using external energy sources can be efficient. 

\subsection{Radiant Energy Sources}
Radiant energy results from the energy transmission in which mass movement does not occur. 
Sunlight, artificial light, radio, and acoustic waves are examples of these. 

\subsubsection{Light Energy Sources}

\begin{figure}[t]
    \centering
    \includegraphics[width=6.3cm]{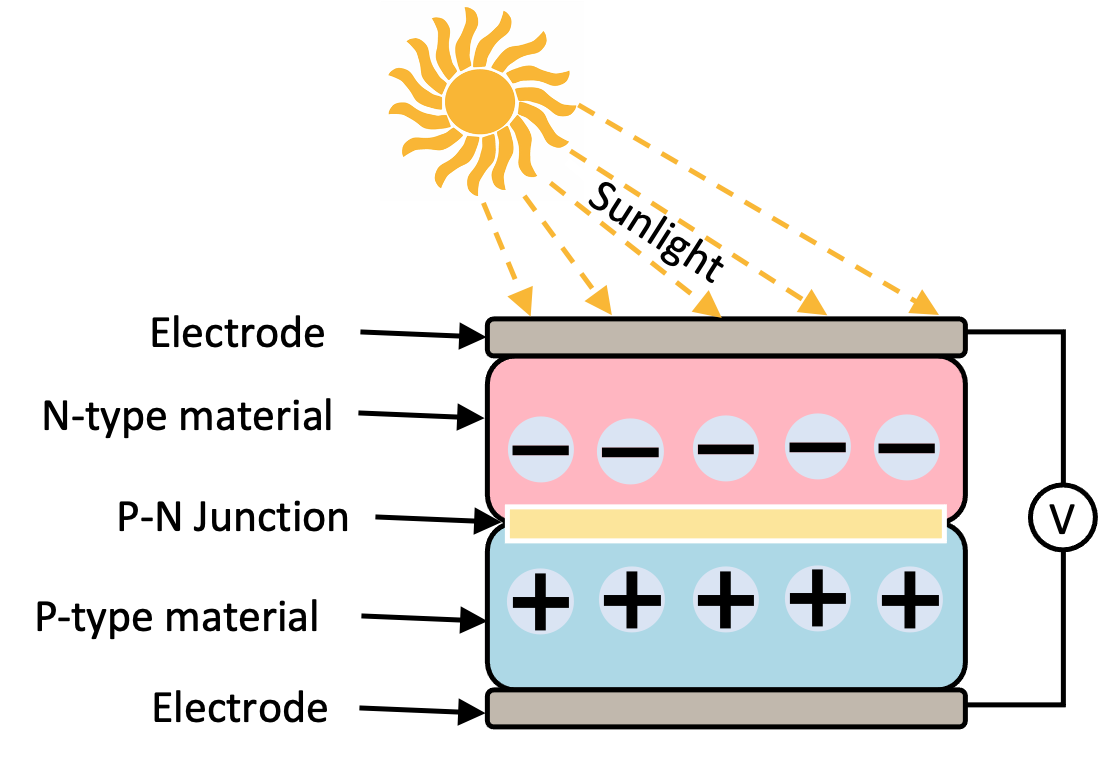}
    \caption{The principle of the photovoltaic effect (adapted from \text{{\color{blue} \cite{mishu2020prospective}}}).}
    \label{workingsolar}
\end{figure}

This section explains the working principle of conventional solar cells, along with the algorithm and different materials that can boost their efficiency. 
Research progress on quantum phenomena-based solar cells is also described and concluded with the techniques to harvest energy from artificial lights.

Sunlight from millions of kilometers away is converted into electrical energy using solar cells concerned with the photovoltaic effect. 
Conventional solar cells have two conductive layers and p and n-type materials \text{{\color{blue} \cite{mishu2020prospective}}}. 
The p and n-type layers consist of highly doped material with holes and electrons, respectively.
According to the photovoltaic effect, photons of sunlight fall onto the p-n junction, and photons push electrons into the n-type layer and holes into the p-type layer shown in Fig. \ref{workingsolar}.
When this solar cell connects with a circuit, electrons leave the n-type region to reach the load. 
Thus, sunlight is converted into electrical energy since the electron flow occurs. 
The obtained output current is described as follows \text{{\color{blue} \cite{mishu2020prospective}}}:
\begin{align}
    I=I_{sc}-I_{0}\left[\exp\left(\frac{V+R_{s}I}{\frac{N_{s}KT}{q}a}\right)-1\right]-\left(\frac{V+R_{s}I}{R_{sh}}         \right),
\end{align}
where the thermal voltage is $V_{t} = \frac{N_{s}KT}{q}$, $I_{sc}$ is the short circuit current, $I_{0}$ is the saturation current, $a$ is the diode ideality constant, $N_{s}$ is the cells connected series, $T$ is the temperature of a p-n junction, $K$ is the Boltzmann constant, which is $K$= 1.38$\times$10$^{-23}$ $J/K$, $q$ is the electron charge, which is $q$= 1.6$\times$10$^{-19}$ $C$, and $R_{s}$ and $R_{sh}$ are series and shunt resistance of a solar cell, respectively. 
However, this equation determines the output current after generating $I_{sc}$ and the generation of $I_{sc}$ directly depends on the area or the size of a solar cell.
Since the intensity of light, temperature, or other environmental conditions are not stable all the time of the day, the output power of solar cells varies \text{{\color{blue} \cite{adu2018energy}}}. 
Therefore, solar cells should operate at the maximum output power level under changing conditions. 
Maximum Power Point Tracking (MPPT) algorithms are proposed in \text{{\color{blue} \cite{wu2016flexible, verma2016maximum}}}. 
Experiments and simulation results demonstrate that the MPPT responds effectively and promptly to load fluctuations. 
A pulse sensor, which can transmit the real-time pulse signal to a smartphone based on a wireless BLE module attached to it, is powered by the MPPT circuit \text{{\color{blue} \cite{wu2016solar}}}. 
Hence, a solar energy harvester with an MPPT circuit can be conveniently used in IoNT to power wearable or portable devices.

Moreover, flexible solar cells are placed on the human body to extract sunlight. 
For instance, a smart wearable bracelet, powered by flexible solar energy harvester increases the operation time of a 200 mAh battery-powered system by 55$\%$ in full mode (4 gesture recognition per 1 second) \text{{\color{blue} \cite{kartsch2018smart}}}. 
Furthermore, a solar energy harvester requires approximately 30 minutes in sunny weather, and 120 minutes in cloudy weather to charge a supercapacitor, the energy-storing module.  
Data acquired by sensors (accelerometer, temperature, and pulse) located on the wrist and chest can be sent to a web-based smartphone application using the BLE module mentioned in \text{{\color{blue} \cite{wu2017autonomous}}} places the device in the taxonomy of IoNT.

Although the mentioned solar cells have flexible structure, they must stay on certain parts (wrist, chest, arm, etc.) of the human body to extract sunlight. 
Hence, it leads to feeling uncomfortable and limits the person's movement. 
Therefore, solar textiles or fiber-based clothes have emerged as a new research topic in recent years \text{{\color{blue} \cite{hatamvand2020recent}}}. 
For instance, the textile can have small crystalline silicon solar cells in its yarns \text{{\color{blue} \cite{satharasinghe2020investigation}}}. 
Other techniques that use textile fabrics as substrates rather than plastic substrates to make the solar fabric more flexible, comfortable, foldable, and durable against bending and stretching consist of dye-sensitized solar cells (DSSCs), organic solar cells (OSCs), and perovskite solar cells (PSCs) \text{{\color{blue} \cite{hatamvand2020recent}}}. 
DSSCs consist of an electrolyte sandwiched from two electrodes, one coated with a semiconductor and dye and the other with platinum or graphite \text{{\color{blue} \cite{ye2015recent}}}. 
Instead of electrolyte (generally liquid), the gap between the electrodes in OSCs consists of an electron transport layer, a hole transport layer, and a photo-active layer composed of a p-type organic semiconductor donor and an n-type organic semiconductor acceptor \text{{\color{blue} \cite{li2018flexible}}}. 
Furthermore, PSCs have a similar structure to OSCs, but this time, the photo-active layer of PSCs consists of crystal structure materials called perovskite \text{{\color{blue} \cite{wang2016stability}}}. 
High power conversion efficiency is achieved by PSCs \text{{\color{blue} \cite{yang2017iodide}}}. 

Moreover, another type of IoHNT exists to harvest energy from sunlight. 
It depends upon the principle of quantum phenomena, called quantum energy harvesting (QEH). 
The name QEH is based on the technique it uses for energy harvesting, not based on its size.
Quantum phenomena-based solar cells use semiconductor dots in solar cells, categorized into metal-semiconductor-based, quantum dot-based, and semiconductor-polymer hybrid cells \text{{\color{blue} \cite{kamat2008quantum}}}. 
The combination of two semiconductors with narrow bandgaps and compatible energy levels optimizes the efficiency of charge separation. 
The alignment of energy levels between the valence and conduction bands of two semiconductors enables the segregation of holes and electrons into distinct entities, leading to the capture of one of the charge carriers at the electrode. 
The most advanced or 4\textsuperscript{th} generation solar cell QEH is quantum dot sensitized solar cells (QDSSCs) \bluecite{sahu2020review}.
Previous generations used quantum dots (QDs) for carrier transportation, whereas QDSSCs use QDs as light-absorbing material.
QDSSC includes a photoelectrode made of a wide bandgap oxide layer like TiO\textsubscript{2}, ZnO, SnO\textsubscript{2}, etc., coated with a QDs layer. 
It consists of a counter electrode, which is a metal or semiconductor electrode with fast kinetics. It also includes an electrolyte with a redox mediator to promote electron transport \bluecite{barcelo2014recent}. 
When these QDs absorb solar radiation as photons, they generate electron-hole pairs, known as excitons. QDSSCs one photonic absorption can lead to more than one exciton generation \bluecite{sambur2010multiple}.
The produced electrons are then quickly injected into the conduction band of the oxide layer and moved toward the external circuit, producing an electric current.
The redox mediator neutralizes the remaining holes in QDs. The counter electrode further reduces the oxidized redox mediator.
The maximum theoretically reported energy conversion efficiency is 44.4\% \bluecite{lee2009highly}.
The most important feature of the quantum phenomena-based IoHNT is that it improves the efficiency of converting energy and provides cost-effective production methods. 
Some detailed discussion on QEH, its material selection, performance, and evidence of enhanced efficiency can be found in \text{{\color{blue} \cite{brown2008quantum,shen2008effect}}}. 
However, a considerable amount of research is required in this domain to achieve significant advancements in QEH.

Besides the Sun, we can also produce light artificially. 
Although the power of artificial light is less than that of the Sun, artificial light can be more favorable in applications where continuity is crucial, especially if energy storage is not feasible. 
Artificial light is an alternative to sunlight and a potential energy source in EH applications. 
In some cases, it is produced as an inevitable byproduct. 
For example, in office buildings, commercial centers, or homes where lighting is necessary for human activities, harvesting energy from artificial light becomes a way to recapture energy that would otherwise be wasted.
Moreover, it can exist in extreme locations, like polar regions, underground environments, etc.
However, since conventional photovoltaic cells are specific for harvesting energy from sunlight, the energy obtained from artificial light using these photovoltaic cells is very low. 
Besides, there is some research to improve the obtained energy from artificial light. 
For instance, dye-sensitized cells (DSC) that achieve an average conversion efficiency of 13.8$\%$ at 1000 lux white LED light is proposed in \text{{\color{blue} \cite{hsiao2020pilot}}}.
Conversely, a design that consists of fluorescent tube light as a source of artificial light and infrared light-emitting diodes (IR-LEDs) is proposed in \text{{\color{blue} \cite{sharma2020indoor}}}. 
IR-LEDs are a special type of LED to emit infrared light. 
Moreover, the IR harvester generates a voltage when it is exposed to light. 
Substantially, this voltage generation happens based on the reverse electroluminescence phenomenon in which semiconductor material emits light when the electrical current or strong electrical field is applied. 
Although a significant amount of power (max 609.9 $\mu$W) from artificial light is suggested in \text{{\color{blue} \cite{sharma2020indoor}}}, the output current is shallow, causing recharge of the supercapacitor or battery within hours. 
Consequently, EH from artificial light is highly suitable for low-power applications. 
A wearable jacket that can be used as an IoHNT, power the vital health monitoring systems (VHMS), and transmit the data obtained from the system to the mobile device or to the doctor via Wi-Fi or BLE module \text{{\color{blue} \cite{khan2022wearable}}}, leading to work as an IoNT module.

\subsubsection{Radio Waves (RF)}
With the increasing applications of self-powered or backup energy storage developed for wearable sensors and implants in the body, IoHNT from RF radiation also become popular in recent years. 
However, the strength of the RF signal decreases when the distance increases. 
\begin{figure} [t]
\centering
{\includegraphics[width=0.6\textwidth]{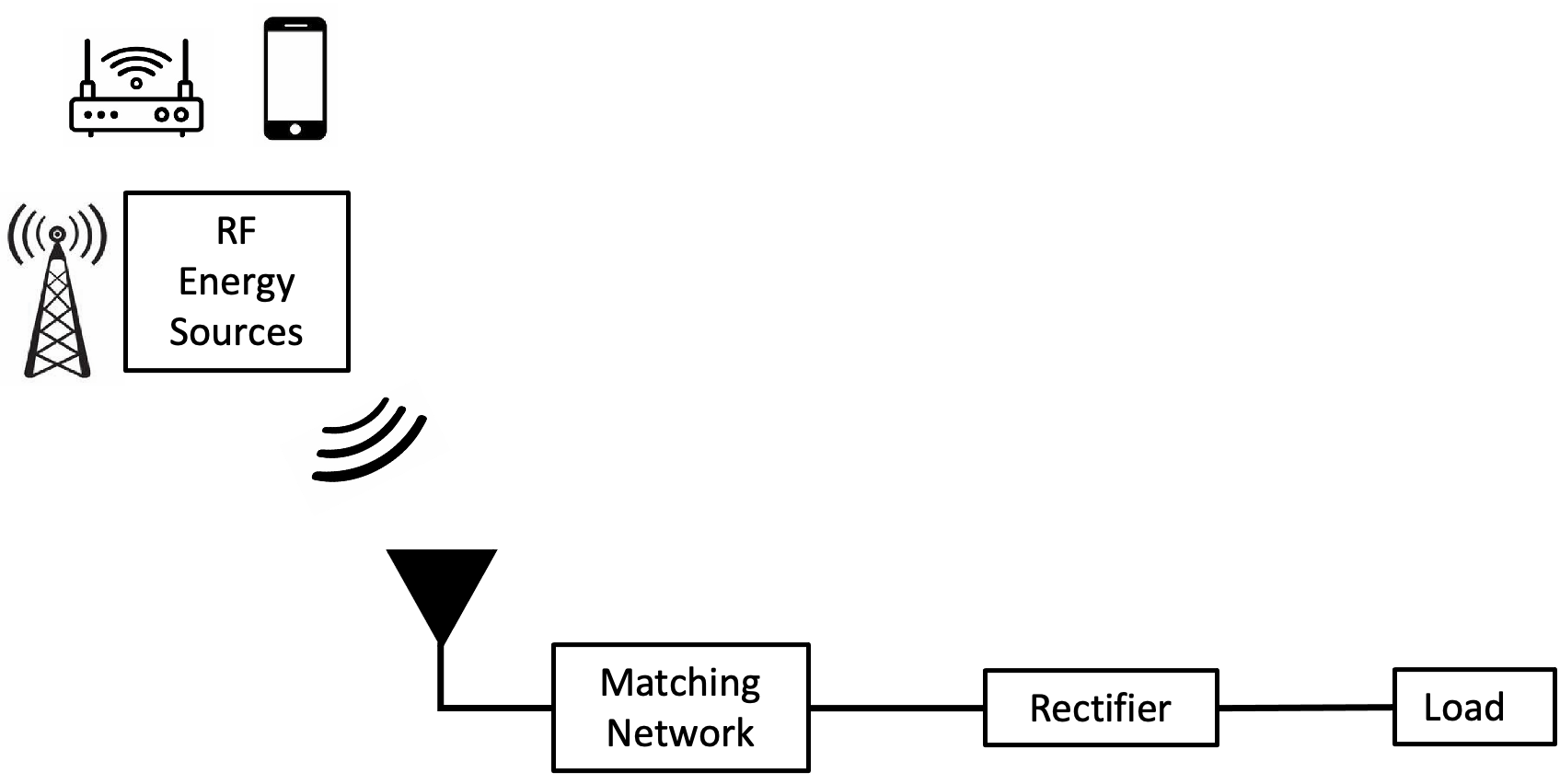}}
\caption{The structure of the rectenna.}
\label{rfstr}
\end{figure}
Thus, a well-designed RF EH system called rectenna, as shown in the Fig. \ref{rfstr}, is crucial which consists of an antenna (receiver), an impedance matching network, and an AC-DC conversion unit (rectifier) \text{{\color{blue} \cite{sidhu2019survey}}}. 
The working principle of rectenna is as follows: The antenna perceives and collects the RF radiations in its environment. 
Then, the antenna transfers collected RF radiations to an impedance-matching network. 
The impedance matching network is crucial in maximizing the RF to DC conversion efficiency in the rectenna by matching the antenna and rectifier impedances. 
Finally, the rectifier converts the AC RF signal into a DC signal. 
The resulting DC signal can directly power low-power devices, or batteries or supercapacitors can store the DC power for later use.
The size of the rectenna depends upon the purpose of use and the operating frequency. 
Typical rectenna's operating frequency depends upon its working wavelength and is classified as microwave, optical, low-powered device \bluecite{donchev2014rectenna}.
However, implementation of the rectennas is based upon the power requirement of IoNTs, as increasing the size of rectennas is directly related to the generated power.

Studies exist on the various rectenna designs to maximize their conversion efficiency. 
For instance, flexible wearable RF energy harvesters for two-way talk radio used in short-range communication without any base station are presented in \text{{\color{blue} \cite{bito2015ambient}}}. 
This RF EH shows a promising maximum output power of 146.9 mW and 43.2 mW with an H-field and E-field harvester. 
Additionally, it is verified for use in a wearable sensing device application. 
Powering an IoNT network may be efficiently achieved using the generated electricity. 
Since RF is not time-dependent like sunlight, it is feasible to operate an IoNT continuously with the assistance of RF EH.
Impedance mismatch causes the received signal to reflect from the rectifier to the antenna, so an efficient RFDC rectifier design is suggested in \text{{\color{blue} \cite{you2020efficient}}} to address this problem. 
On the other hand, in recent years, studies have been conducted on textile-based RF EH to increase durability, flexibility, and comfortability. 
A rectenna that is pasted on the human body can harvest energy from the signal of a Wi-Fi router, proposed in \text{{\color{blue} \cite{lin2018wearable}}}. 
A 2$\times$2 array rectenna can produce a DC voltage of 1.05 V when the distance between the router and the rectenna is 1.5 meters, potentially leading to use in healthcare applications of IoNT \text{{\color{blue} \cite{lin2018wearable}}}. 
A jacket integrated with textile-based rectennas is proposed in \text{{\color{blue} \cite{vital2019textile}}} to harvest energy from Wi-Fi signals in the environment. 
A 2$\times$3 array can produce an average DC power up to 80 $\mu$W at a 60 cm distance from the source.

Lately, fifth-generation (5G) cellular network technology and beyond offers a high-frequency millimeter range to increase service quality and data rate for users \text{{\color{blue} \cite{al2019millimetre, chen2020massive}}}. 
Since the high-frequency millimeter region includes a large amount of RF power, a rectenna operating in that region has great potential to obtain high DC power. 
In this regard, a rectenna that consists of a graphene field-effect transistor (FET)-based rectifier for a 29-46 GHz millimeter-wave band is elaborated in \text{{\color{blue} \cite{singh2020compact}}}. 
The rectenna achieves a maximum conversion efficiency of 80.32\% when subjected to an input power of 2 dBm and a gain level of up to 8.12 dBi. 
The rectenna being considered has a frequency range of 17 GHz and produces an output voltage of 6.38 V \text{{\color{blue} \cite{singh2020compact}}}. 
The compact dimensions of this rectenna, together with its output characteristics, render it suitable for powering IoNT.
An RF EH has the ability to catch, gather, transform, and retain the ambient radio frequency energy, which may then be transmitted to IoNTs wirelessly.
It offers a durable and almost infinite energy supply for IoNTs, hence significantly prolonging their lifespan and ensuring uninterrupted functioning.

\subsubsection{Acoustic (Sound) Waves}
\begin{figure}[t!]
\centering
{\includegraphics[width=0.45\textwidth]{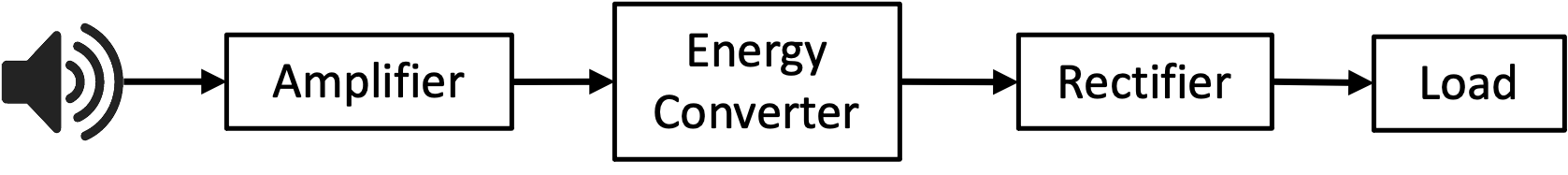}}
\caption{The conversion of the acoustic energy into electrical energy.}
\label{acousticstr}
\end{figure}

Acoustic EH (AEH) is not a good option for high-power-consuming applications due to its low power density \text{{\color{blue} \cite{choi2019brief}}}. 
However, it can meet the power demand of low-power applications, including nanodevices \text{{\color{blue} \cite{wang2020powering}}}. 
The structural design of a proper harvester is crucial to get as high as possible energy from sound waves \text{{\color{blue} \cite{yuan2019recent}}}. 
Therefore, in the literature, various structural designs of AEHs are used to overcome the low power density. 
Operating at resonance frequency is another crucial property to maximize the output power of the harvester used to amplify the low-amplitude sound waves coming from the environment \text{{\color{blue} \cite{yuan2019recent}}}. 
There are several stages in AEH to convert acoustic energy into electrical energy, as shown in Fig. \ref{acousticstr}. 
The first stage amplifies the sound waves using a resonator or an acoustic metamaterial \text{{\color{blue} \cite{choi2019brief}}}. 
The second stage is energy conversion realized by piezoelectric, electromagnetic, or triboelectric effects \text{{\color{blue} \cite{choi2019brief}}}. 
The third stage includes rectification (AC-DC conversion) and power management circuits \text{{\color{blue} \cite{choi2019brief}}}. 
Finally, obtained electrical energy can be stored by a supercapacitor and battery or directly used for applications \text{{\color{blue} \cite{pillai2014review}}}. 
An advantage of employing acoustic energy harvesting in the realm of IoNT is the widespread availability of sound sources that may be harnessed, making it a virtually boundless power source, depending on the surroundings. 
Biomedical applications can utilize electricity generated from internal bodily noises, such as pulses or breathing sounds. 

\begin{figure}[t]
    \centering
    \includegraphics[width=0.45\textwidth]{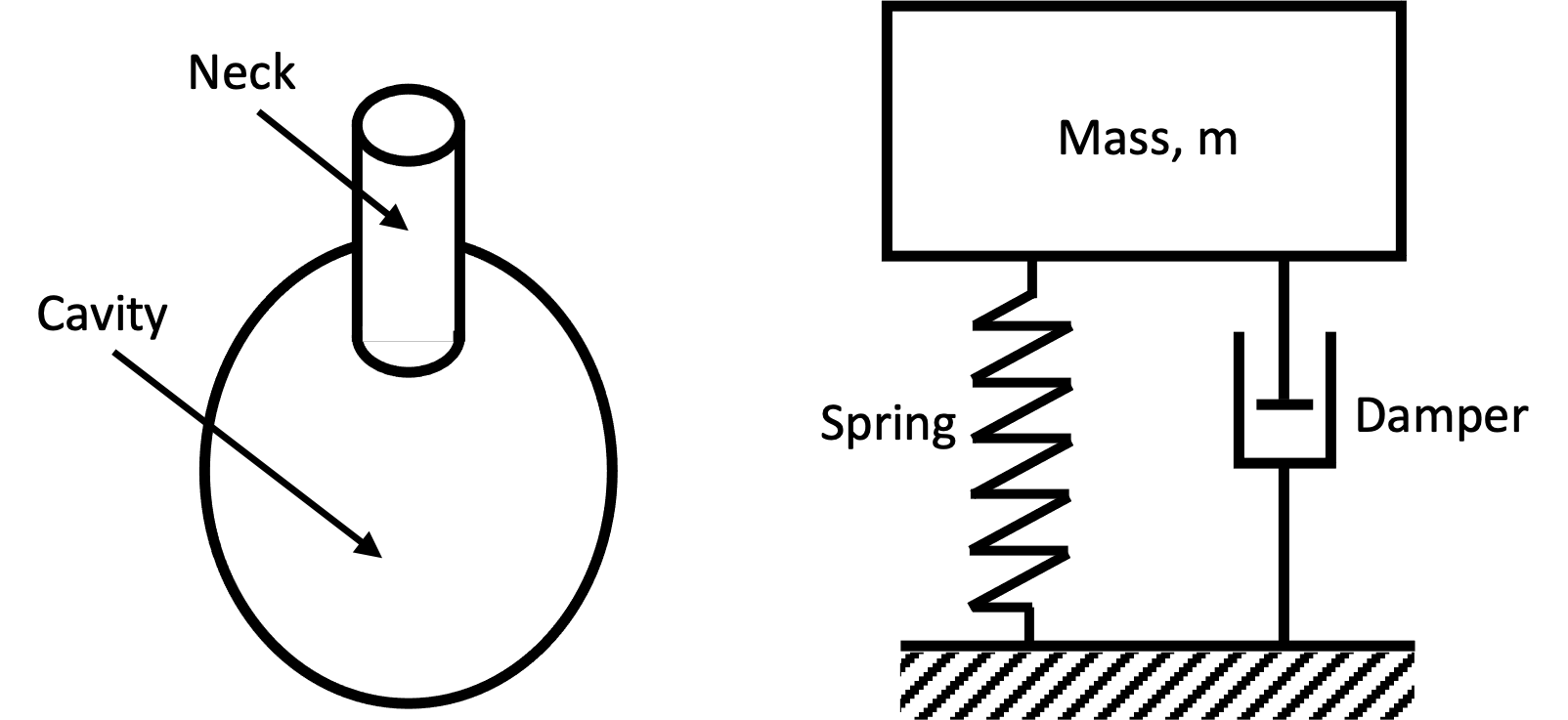}
    \caption{Helmholtz resonator and its lumped-element model.}
    \label{acousticapp1}
\end{figure}

\begin{figure}[t!]
    \centering
    \subfloat[]{\includegraphics[width=0.44\textwidth]{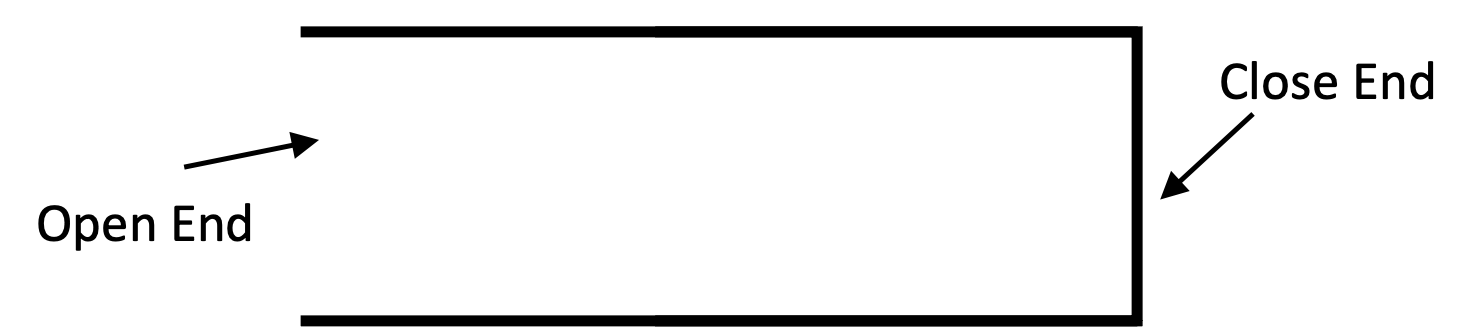}}\hspace{0.1mm}
    \subfloat[]{\includegraphics[width=0.44\textwidth]{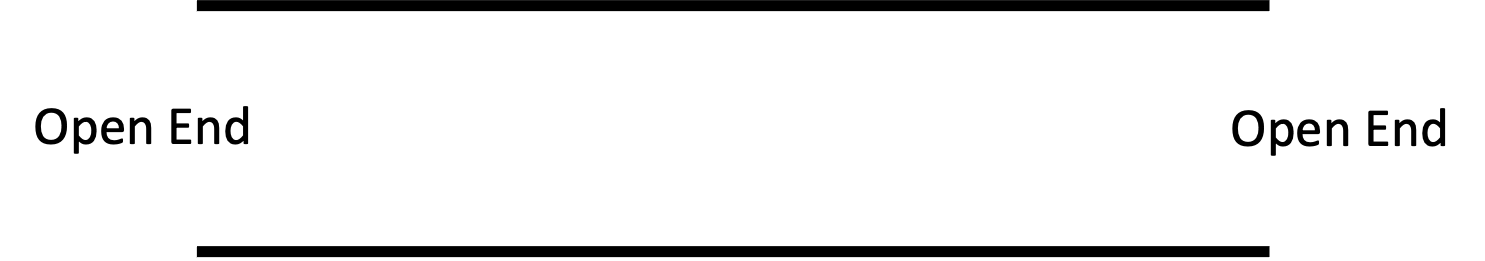}}
    \caption{Schematics of resonators. (a) Quarter wavelength resonator. (b) Half wavelength resonator.}
    \label{acousticapp2}
\end{figure}

Moreover, for the proper working of applications, the obtained energy from the harvester must be continuous at a certain level. 
Therefore, MPPT for acoustic energy harvester is another critical point \text{{\color{blue} \cite{gu2021research}}}. 
For a load of 110 k$\Omega$ this AEH with MPPT produces  105.1 $\mu$W of power.
A conventional AEH generally consists of a resonator, membrane, and piezoelectric material \text{{\color{blue} \cite{salem2020acoustic}}}. 
The resonator operates at the resonant frequency, ensuring the incoming sound waves have a high amplitude \text{{\color{blue} \cite{yuan2019recent}}}. 
There are three types of resonators, including Helmholtz, as shown in Fig. \ref{acousticapp1}, quarter-wavelength, and half-wavelength resonators \text{{\color{blue} \cite{choi2019brief}}}, as shown in Fig. \ref{acousticapp2}. 
A Helmholtz resonator consists of a cavity connected with a neck \text{{\color{blue} \cite{yuan2019recent}}}. 
The sound pressure rises from neck to cavity and spreads inside the cavity uniformly at the resonant frequency \text{{\color{blue} \cite{yuan2019recent}}}. 
Another resonator type is a quarter-wavelength tube resonator with open and closed ends used to amplify sound waves \text{{\color{blue} \cite{yuan2019recent}}}. 
As the name suggests,  the length of the tube is equal to the quarter wavelength of the sound waves \text{{\color{blue} \cite{yuan2019recent}}}. 
The working mechanism of the quarter-wavelength resonator is similar to that of the Helmholtz resonator, with the open and closed ends acting as the neck and cavity, respectively \text{{\color{blue} \cite{choi2019brief}}}. 
The half-wavelength resonator, which has both open ends, is also used to amplify the sound waves  \text{{\color{blue} \cite{choi2019brief}}}. 
The resonant frequency of the quarter-wavelength and half-wavelength resonator can be given as follows:

\begin{align}
    f&=\frac{v}{4L} \ \ \ \text{(quarter wavelength resonator)},\\[6 pt]
    f&=\frac{v}{2L} \ \ \ \text{(half wavelength resonator)},
\end{align}
where the $v$ is the speed of sound, and the $L$ is the length of the tube.

Ways to harvest acoustic energy do not only rely on resonators. 
For instance, a piezoelectric nanogenerator consisting of PVDF nanofiber filled with titanium dioxide (TiO$_{2}$) nanoparticles to harvest energy from human motion and sound waves is proposed in \text{{\color{blue} \cite{alam2018biomechanical}}}. 
The produced output power of this nanogenerator is 4 $\mu$W with an efficiency of 61\%. 
A paper-based triboelectric nanogenerator whose working mechanism depends on contact separation mode to obtain energy from human talking is suggested in \text{{\color{blue} \cite{fan2015ultrathin}}}, can generate a volume power density of up to 121 mW/m$^2$ when subjected to a sound pressure level of 117 dB. 
Even though it seems feasible to wirelessly power an IoNT or a nanodevice, considering the technology and the produced output power, it is still an open-ended research domain.

On the other hand, metamaterials instead of resonators can also provide sound pressure amplification \text{{\color{blue} \cite{yuan2019recent}}}. 
Metamaterials are unusual materials and have exceptional properties not observed in nature, being produced in a laboratory environment \text{{\color{blue} \cite{choi2019brief}}}. 
They also have the ability of sound-wave manipulation, including amplification, which is quite essential for the AEH \text{{\color{blue} \cite{yuan2019recent}}}.
Conversion of sound energy utilizing piezoelectric technology by the acoustic metamaterial plate (meta-plate) is proposed and investigated in \text{{\color{blue} \cite{xiao2023metamaterial}}}. 
At 30 Pa sound pressures, it is theoretically capable of harvesting 3.09 mW of power with an optimal load resistance of 28180 $\Omega$.
Using metamaterials coupled with a resonator has higher output power (93.13 $\mu$W at 10 Pa sound pressure) \text{{\color{blue} \cite{ma2021metamaterial}}} than using only one metamaterial plate ($\sim$7 $\mu$W at 10 Pa sound pressure) \text{{\color{blue} \cite{ma2020acoustic}}} in AEH. 
Therefore, using metamaterials in AEH may increase the output power. 
The use of metamaterials in acoustic IoHNT for powering IoNT devices provides numerous advantages, including increased efficiency, selective operation at specific frequencies, improved sensitivity, the ability to capture a wide range of frequencies, adjustability, adaptability, and a low environmental impact. 
The combination of these benefits results in the development of IoNT devices that are more efficient and adaptive, with the ability to harvest energy from the surrounding acoustic environment.

\subsection{Flow-based Energy Sources}
Flow energy results from the energy transmission in which mass movement does occur. In other words, these sources transmit energy by moving the mass particles.

\subsubsection{Water}
Water is an essential liquid for the maintenance of human life, and water motion occurs naturally as well. 
Therefore, water motion can potentially be a low-power application in EH systems, especially intra-body devices. An EH method from water drops is proposed in \text{{\color{blue} \cite{kwon2014effective}}}. 
Substantially, when charge-neutral water droplets fall on a dielectric material, they charge. 
A thin electric double layer (EDL) occurs between the water and the dielectric material, providing a charge current to the electrode under the dielectric layer. 
It provides a peak voltage of $\sim$3.1 V and a peak current of $\sim$5.3 $\mu$A \text{{\color{blue} \cite{kwon2014effective}}}. 
This approach can be a base for future-generation wearable EH systems, such as a jacket that harvests energy from raindrops. 
A droplet energy harvesting (DEH) panel, which scavenges energy from multi-position droplet impacts, is proposed in \text{{\color{blue} \cite{D2EE00357K}}}. 
It is a cross-talk-free, fully transparent DEH panel that provides high output performance. 
Hence, harnessing water energy is advantageous in locations with a consistent water flow, such as industrial and agricultural settings. 
Ecosystems (rivers, oceans, etc.) and their properties may be tracked, i.e., environmental monitoring can be carried out with this IoHNT approach coupled with an IoNT device. 

\subsubsection{Wind}
The wind is another ambient energy source that can potentially be used in EH systems. 
However, conventional wind energy harvesters called windmills are unsuitable for IoNTs EH systems. 
Therefore, through the advancement of micro/nano-manufacturing technology, small-scale wind energy harvesters have started to be manufactured \text{{\color{blue} \cite{fu2020overview}}}. 
Additionally, piezoelectric, electrostatic, and triboelectric EHs can convert wind to electrical energy for small-scale devices. 
For instance, a piezoelectric wind energy harvester that consists of a cantilever beam and a square prism attached to the end of the beam is proposed in \text{{\color{blue} \cite{hu2019performance}}}. 
When two identical harvesters are arranged in tandem and placed in a tunnel, energy is obtained from the relative motions of harvesters with the effect of low turbulence flow in the tunnel \text{{\color{blue} \cite{hu2019performance}}}. 
Moreover, triboelectric wind energy harvesters that consist of a stator and a rotor are suggested in \text{{\color{blue} \cite{bi2020optimization}}}. 
A wind turbine induces the relative motion between the stator and rotator so that the charges are transferred between the electrodes, and energy is harvested according to triboelectric phenomena \text{{\color{blue} \cite{bi2020optimization}}}. 
A new wind-based energy-harvesting technology using magnetoelastic generators (MEG), based on the principle of electromagnetic induction, is proposed in \text{{\color{blue} \cite{zhao2022soft}}}.
Under natural wind conditions, MEG has a power density of 0.82 mW/cm$^{2}$ at the wind speed of 20 m/sec \text{{\color{blue} \cite{zhao2022soft}}}.

\subsection{Influence of external energy sources based IoHNT on IoNT communication}
External energy sources, such as sunlight, artificial light, RF, acoustic, water, and wind, can be effectively harvested by IoHNTs. These sources are crucial for powering IoNT devices, especially those deployed in outdoor or well-lit environments.
IoHNTs equipped with photovoltaic or piezoelectric materials can convert external energy into electrical power, ensuring continuous power supply wirelessly. 
It is critical for driving the nanoscale communication components, such as transceivers and antennas, essential for establishing and maintaining communication links in IoNT networks. 
The energy harvested can support the operation of IoNTs' mesh networks (elaborated in Sec. \ref{mesh}), where nodes must communicate across multiple hops to relay data effectively.
A stable and continuous power supply from external sources-based IoHNTs allows IoNTs to support longer transmission ranges and maintain robust signaling.
This capability is significant for IoNT-specific communication protocols, such as IEEE 802.15.4 (elaborated in Sec. \ref{par:Communication Protocols}), designed to optimize energy usage and reduce latency in nanoscale networks. 

External energy harvesting facilitates the large-scale deployment of IoNT devices, allowing them to integrate seamlessly into broader IoT networks. 
These devices can serve as nodes in extensive IoT systems, contributing to real-time data collection and analysis. 
The interoperability of IoNTs with existing wireless communication standards, such as 5G and upcoming 6G networks, ensures the expansion of IoT ecosystems.
As nanotechnology advances, IoNT networks will become more autonomous and capable of sustaining long-term operations in diverse environments. 
Future IoNT applications could include intelligent agriculture systems, where nanoscale sensors powered by external energy sources based IoHNTs continuously monitor soil conditions and transmit data to central hubs. 
Integrating IoNT with advanced networking protocols and edge computing will further enhance the scalability and efficiency of these networks.

As the energy that can be obtained from environmental sources is far superior to the energy generated within the human body, obtaining electrical power through these sources can be pretty beneficial. 
IoNTs need continuous power supply for continuous operation and wirelessly transmit the data to the gateway.
Hence, depending on the condition and application, the body-centric IoHNT can be more favorable to power IoNT systems in some conditions.

\section{Body-centric Energy Sources and Harvesting}
\label{sec:Body-centric Energy Sources and Harvesting}
One of the most significant obstacles in fully leveraging the capabilities of IoNT, particularly in the realm of biomedical applications, pertains to the efficient and sustainable energizing of nanodevices. 
Harvesting energy internally in the body enables the uninterrupted and prolonged functioning of IoNT devices. 
This is essential for applications like real-time health monitoring, precise medicine administration, and improved diagnostics.
We consider \textit{Body} as the human body, and sources are classified according to that. 
However, the applicability of produced IoHNT extends to non-human bodies, such as the environment and vehicles.
Body-centric energy sources are the types of energy originating from living organisms directly or emerging inside them, e.g., body movements and biofuel cells.

\subsection{Human Body Energy Sources}
In the discussion, we classify the human body energy sources as the body movements, body heat, and biofuel cell energy sources.
\subsubsection{Body Movements}
Most body parts, e.g., hand, arm, leg, elbow, neck, and foot, become active when a person starts running, walking, or exercising from a resting state, which refer to kinetic energy in physics. 
Triboelectric, piezoelectric, electrostatic, and electromagnetic induction principles provide the ability to convert this kinetic energy into electrical energy.

\paragraph{Triboelectric Effect} 
The energy is harvested by two materials, one having a tendency to gain and the other one having a tendency to lose an electron. 
The harvesting occurs by the coupling effects between the triboelectric effect and electrostatic induction emerging from separation or sliding. 
A way to utilize the body to harvest energy is by using triboelectric materials, which can convert biomechanical energy from the human body into electrical energy with the triboelectrification effect. 
Since the triboelectric effect is quite prevalent, triboelectric nanogenerators (TENG) stand out as one of the best choices to harvest energy. 
Numerous materials exist that can be used as TENGs.  
There is an increasing attraction towards TENGs by researchers \text{{\color{blue} \cite{tri_hea, body_tri}}}, especially in motion monitoring and related healthcare fields. 
Different moving parts of the body can be used as an energy source via the conjunction of triboelectrification and electrostatic induction, such as walking \text{{\color{blue} \cite{gu2017antibacterial}}}, eye blinking \text{{\color{blue} \cite{eye}}}, finger movement \text{{\color{blue} \cite{yan2020linear}}}, breathing and many other different activities \text{{\color{blue} \cite{canan}}}.
A brief list for activities is given in Table \ref{table:mert4}, summarized from \text{{\color{blue} \cite{riemer2011biomechanical, niu2004evaluation, park2016micro}}}.

\begin{table}[t!]
   \centering
    \caption{Human activities for triboelectric EH}
    \label{table:mert4}
    
    \begin{tabular}{ |>{\centering\arraybackslash}p{1cm} |>{\centering\arraybackslash}m{2.5cm} |>{\centering\arraybackslash}m{1.6cm} |>{\centering\arraybackslash}m{3cm} |}
    \hline
    \centering
    
    \textbf{Motion} & \centering  \textbf{Motion Properties} & \centering \textbf{Output Power} $(W)$ & \textbf{TENG Material Properties}\\ \hline     
    
    Elbow & high frequency, \centering high amplitude & \centering $2.1$ & flexible,  breathable, skin-friendly \\ \hline
    Finger &  high frequency, \centering low amplitude & \centering $0.23-0.4$ & flexible, breathable, skin-friendly, soft\\ \hline
    Chest & high frequency, \centering low amplitude & \centering $-$ & flexible, breathable, skin-friendly \\ \hline
    Leg & high frequency, \centering low amplitude & \centering $2-20$ & durable, breathable\\
    \hline
    \end{tabular}
\vspace{0.7mm} 
\end{table}

At first, the TENGs were placed directly in contact with the body parts, which not only caused a decrease in the robustness of the TENGs but also corrodes the material itself due to friction and body heat, as exemplified in \text{{\color{blue} \cite{glove}}}. 
However,  recent studies \text{{\color{blue} \cite{non, no_con}}} suggest that harvesting energy with triboelectrification is possible without suffering direct contact consequences.
Therefore, TENG energy harvesters can appear in many forms through direct contact and non-direct contact, attached to fabric, cloth, glasses, and shoes according to the desired input source area. 
TENGs are used vastly in our daily lives to power wearable biomedical sensors, such as cardiac and pneumology sensors \text{{\color{blue} \cite{TAT2021112714}}}. 
A wood-derived, durable, biocompatible, and with antifungal activity wearable TENG is fabricated in \text{{\color{blue} \cite{park2022biocompatible}}}. 
It can produce a maximum output voltage of 80 V and is stable even at 100,000 cycles.

Electrostatic induction is a well-known technique to generate static electricity. 
When any two materials that have different charges on their surfaces come close to each other, redistribution and transfer of the charges occur. 
Similar to that sense, the triboelectric effect allows charges to transfer between any two materials when they contact each other \text{{\color{blue} \cite{wang2017maxwell}}}.  
A simple TENG consists of two triboelectric layers that are charged oppositely, an air gap between them, and two metallic electrodes \text{{\color{blue} \cite{zhou2020engineering}}}. 
When these two triboelectric layers make contact or slide motion, the current is generated by the flow of induced charges and the potential difference that occurs between the electrodes. 
The triboelectric potential $(V_T)$ is expressed as follows \text{{\color{blue}
\cite{mishu2020prospective}}}: 
\begin{align}
    V_{T} &= -\frac{\rho_{T}d}{\varepsilon_{0}} ,
\end{align}
where $\rho_{T}$ is the triboelectric charge density, $\varepsilon_{0}$ is the vacuum permittivity, $d$ is the gap between two triboelectric layers.
The current ($I_T$) can be given by \text{{\color{blue}
\cite{mishu2020prospective}}}:
\begin{align}
    I_{T} = \frac{\partial}{\partial t}(C_TV_T) &= C_T\frac{\partial V_T}{\partial t} + V_T\frac{\partial C_T}{\partial t} ,
\end{align}
where $I_{T}$ is the triboelectric current, $C_{T}$ is the capacitance, and $V_{T}$ is the voltage.

\begin{figure*}[t!]
    \centering
    \begin{subfigure}{0.24\textwidth} 
        \centering
        \includegraphics[width=\textwidth]{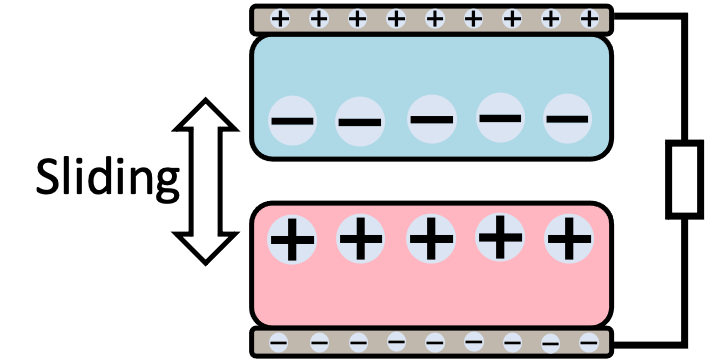} 
        \caption{}
    \end{subfigure}
    \hfill 
    \begin{subfigure}{0.22\textwidth} 
        \centering
        \includegraphics[width=\textwidth]{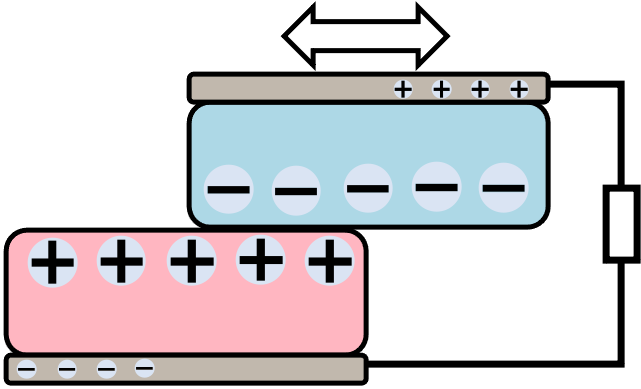} 
        \caption{}
    \end{subfigure}
    \hfill 
    \begin{subfigure}{0.22\textwidth} 
        \centering
        \includegraphics[width=\textwidth]{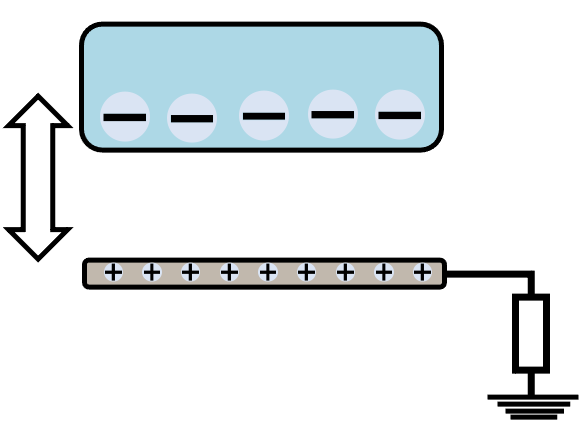} 
        \caption{}
    \end{subfigure}
    \hfill 
    \begin{subfigure}{0.24\textwidth} 
        \centering
        \includegraphics[width=\textwidth]{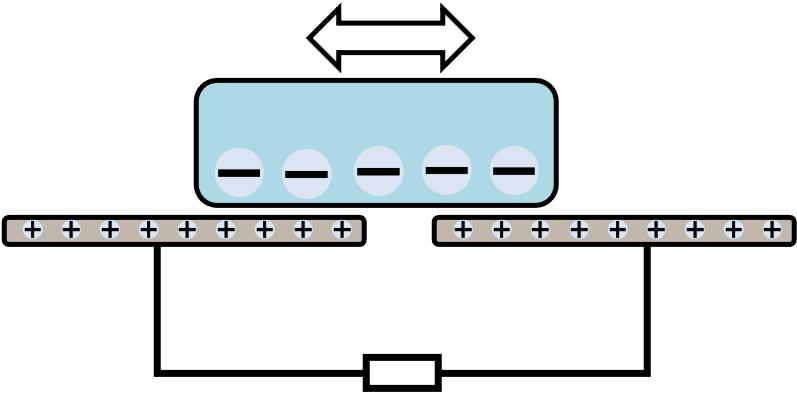} 
        \caption{}
    \end{subfigure}
   \caption{Four main working mechanisms of the TENG. (a) The working principle of the contact-separation mode. (b) The working principle of the lateral sliding mode. (c) The working principle of the single electrode mode. (d) The working principle of the free-standing mode.}
    \label{4TENG}
\end{figure*}

The TENG has four main operation modes: vertical contact separation, lateral sliding, single-electrode, and free-standing bioelectric layer \text{{\color{blue} \cite{wang2015triboelectric}}}, as shown in Fig. \ref{4TENG}. 
TENG in vertical contact-separation mode, as shown in Fig. \ref{4TENG}a, can sense any direct forces, such as impact, pressing, or shock, while in lateral sliding mode, it can sense any sliding motion that occurs by running or jumping, as shown in Fig. \ref{4TENG}b. 
The single-electrode mode has a better practical design than the other modes due to not connecting the triboelectric layer to any external circuit, meaning it moves freely on the electrode, as shown in Fig. \ref{4TENG}c. 
This mode can be used to harvest energy from human movements like walking and finger movements \text{{\color{blue} \cite{wang2015triboelectric}}}. 
The free-standing mode can harvest energy from moving objects, but the whole system is mobile, not connected to the ground \text{{\color{blue} \cite{wang2021triboelectric}}}, as depicted in Fig. \ref{4TENG}d.

\begin{figure}[t]
    \centering
    \includegraphics[width=5.7cm]{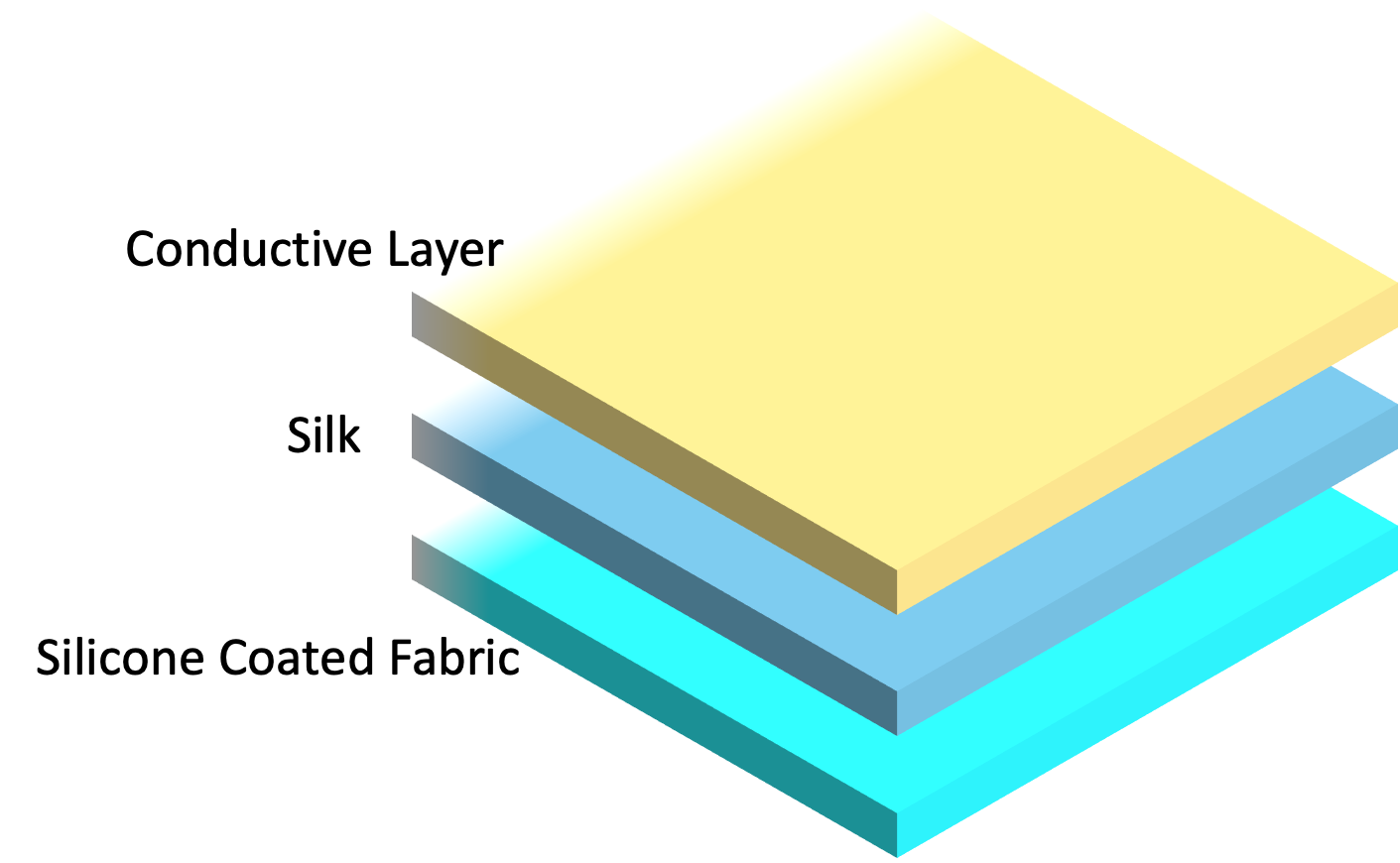}
    \caption{Textile TENG device structure (adapted from \text{{\color{blue} \cite{cao2022full}}}).}
    \label{output_sound}
\end{figure}

TENGs can also be attached to clothes to generate energy from the movement. 
TENG with polytetrafluoroethylene (PTFE) and copper as tribo-pairs working in free-standing provides EH from the back and forth arm movement during the running exercise \text{{\color{blue} \cite{Songeaay9842}}}. 
Textile TENG, which consists of silk fabric and hollow fiber pumped with gallium-based liquid metal (not harmful to the human body), uses the vertical contact-separation mode to convert human motion into electrical energy \text{{\color{blue} \cite{WANG2020104605}}}.  
Weft yarns harvest energy from motion using lateral-sliding mode. 
An entire textile TENG to generate energy from sound, motion, and wind is developed in \text{{\color{blue} \cite{cao2022full}}}, as shown in Fig. \ref{output_sound}. 
If the sound conversion efficiency is considered, it can convert low-frequency sound (e.g., 75 Hz) into a maximum instantaneous power density of 19.4 mW/m$^2$. 
A multi-mode stretchable and wearable TENG (msw-TENG), which operates in contact separation mode is developed in \text{{\color{blue} \cite{wu2022multi}}}.  
Furthermore, some wearable TENG-based EHs are described and reviewed in detail in \text{{\color{blue} \cite{wang2021design}}}. 
Due to size, lifespan, and sustainability concerns, conventional battery technology is frequently inconceivable for nanodevices, making use of IoHNT for IoNTs. 
TENG-based IoHNT presents the potential to power IoNTs seamlessly, making them capable of sustained operation.
This integration is crucial for applications in which sustaining power supply is difficult or other IoHNT techniques are inefficient to implement, such as intelligent textiles and medical devices operating in the proximity of the human body.

\paragraph{Piezoelectric Effect}
The piezoelectric effect is the ability to generate a surface electrical charge upon applying force to a piezoelectric material.    
It is a two-sided phenomenon, with a direct and converse piezoelectric effect, which is depicted in Fig. \ref{fig:pie}. 
Converse is the ability to induce mechanical stress from exposure to an electric field. 
The relation between the direct and converse piezoelectric effect is given in \text{{\color{blue} \cite{erturk2011piezoelectric}}} and can be expressed as:

\begin{figure}[t!]
\centering
	\hspace{0.1mm}
        \begin{subfigure}[b]{0.24\textwidth}
		{\includegraphics[width=\textwidth]{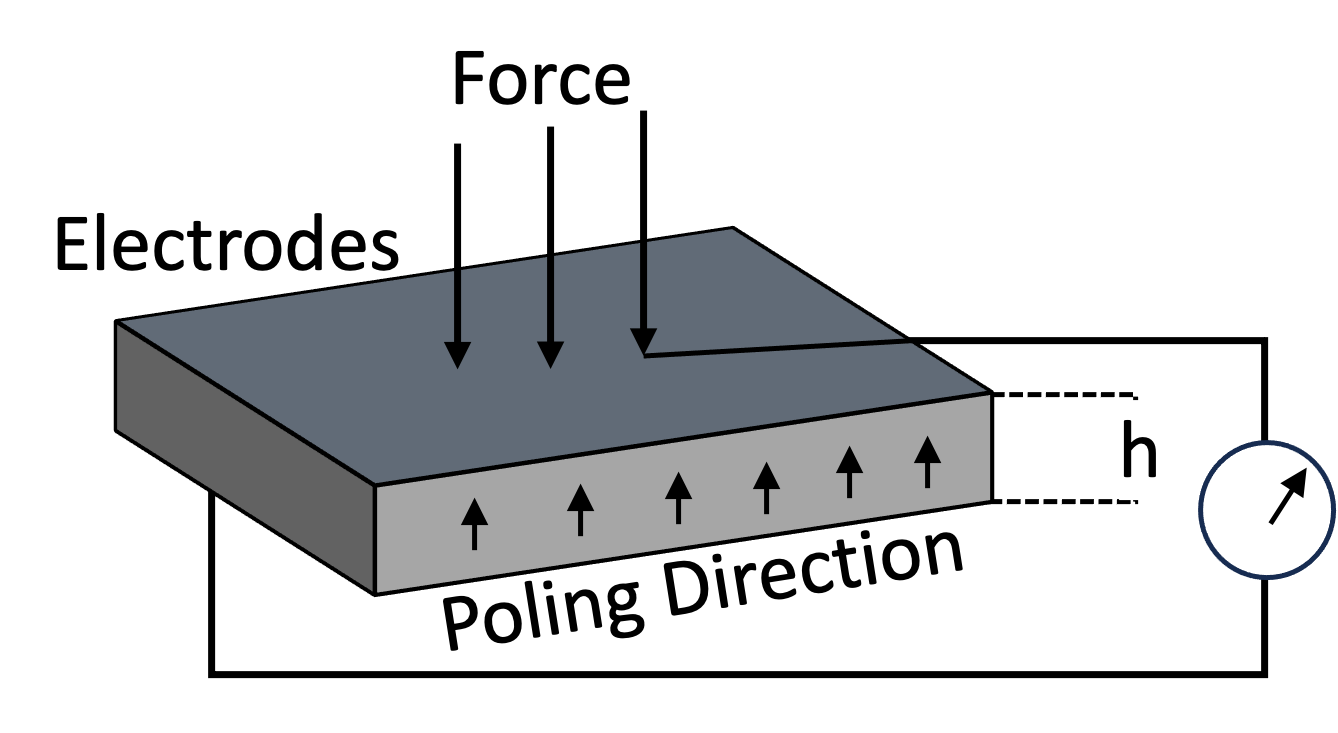}}
		\caption{}
	\end{subfigure}
	\hspace{0.1mm}
		\begin{subfigure}[b]{0.22\textwidth}
		{\includegraphics[width=\textwidth]{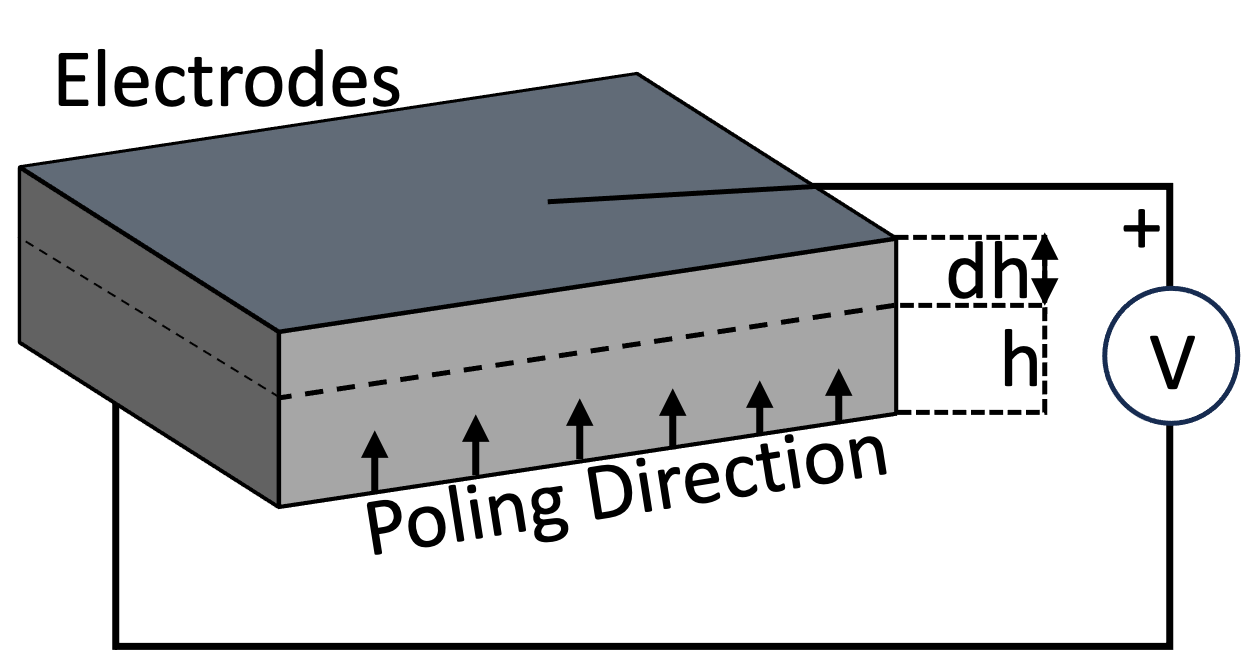}}
		\caption{}
	\end{subfigure}
	\caption{Piezoelectric Phenomenons (adapted from \text{{\color{blue} \cite{devasia2007survey}}}). (a) Direct Effect. (b) Converse Effect.}
    \label{fig:pie}
\end{figure}  
    
\begin{equation}
\label{eqn:matrix}
\begin{bmatrix}
\delta \\
D
\end{bmatrix}
=
\begin{bmatrix}
s^E & d^t\\ d & \varepsilon^T
\end{bmatrix}
\begin{bmatrix}
\sigma \\  E
\end{bmatrix},
\end{equation} 
where $\delta$, $\sigma$, $E$, $s$, $d$, $D$, $\varepsilon$ refer to strain, stress, electric field, elastic compliance, piezoelectric coefficient, electric displacement, and the dielectric constant, respectively, whereas the superscripts $E$ and $T$ denote that the respective constants are evaluated at the constant electric field and constant stress, respectively. 
The superscript $t$ refers to transpose \text{{\color{blue} \cite{doi:10.1063/1.5074184}}}.
    
\begin{figure}
    \centering
    \includegraphics[width=6cm]{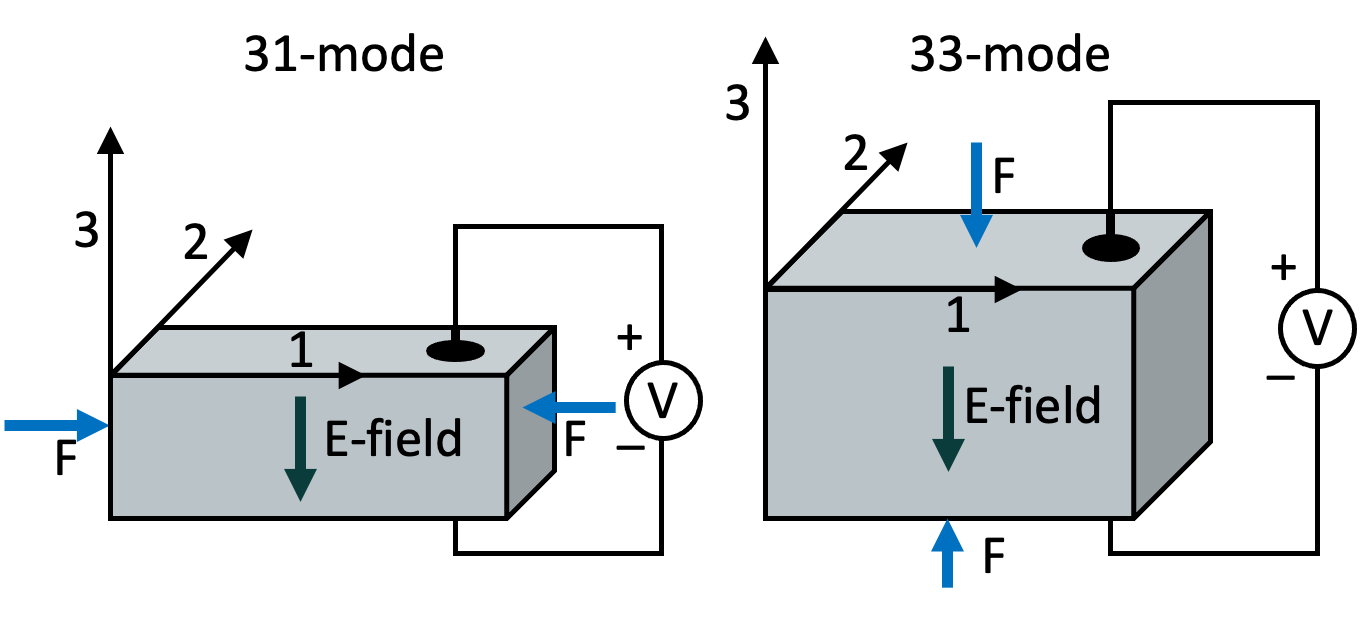}
    \caption{Working principle of 31 and 33-mode piezoelectricity.}
    \label{31-33}
\end{figure}
Different configurations of piezoelectric-based circuitries exist, such as 31-mode and 33-mode. 
31 mode is when an electric field is applied in the 1-axis direction, and the mechanical stress or strain is applied along the 3-axis. This mode is typically favored in a thin film or plate-like material. On the other hand, in 33 mode, both the electric field and the mechanical deformation are applied along the same axis (3-axis). This mode is used in bulk piezoelectric material. These two modes are illustrated in Fig. \ref{31-33}.
Some more combinations are 31-mode bimorph cantilever in series and parallel connections and 31-mode and 33-mode unimorph cantilever configurations. Configurations' performances differ according to the application area, such as in MEMS implementations; an unimorph cantilever configuration is generally chosen \text{{\color{blue} \cite{doi:10.1063/1.5074184}}}. 
Other than the configuration type, other intrinsic and extrinsic factors influence piezoelectric EH's performances. 
The open circuit voltage of the piezoelectric layer is calculated as follows:
\begin{equation}
\label{eqn:voc}
    V_{oc} = \frac{d_{ij}\sigma_{ij}g_e}{\varepsilon} ,
\end{equation}
\textcolor{blue}{(\ref{eqn:voc})} shows that the open circuit voltage $V_{oc}$ has a linear relationship between the piezoelectric coefficient ($d_{ij}$), applied stress ($\sigma_{ij}$), and the gap distance between two electrodes ($g_{e}$), respectively. 
However, it is disproportionate to the $\varepsilon$, which is the permittivity of the substance. The details regarding the configurations of the piezoelectric EH are inquired in \text{{\color{blue} \cite{doi:10.1063/1.5074184}}}. \par 

\begin{table}[t!]
    \centering
    \begin{center}
    \caption{Short circuit current ($I_{sc}$) and open circuit voltage ($V_{oc}$) values of commonly used piezoelectric nanogenerators}
    \label{table:mert_pie}
    \centering
    \begin{tabular}{ |>{\centering\arraybackslash}m{2cm} |>{\centering\arraybackslash}m{1.3cm} |>{\centering\arraybackslash}m{1.3cm} |>{\centering\arraybackslash}m{0.7cm} |}
    \hline
    \centering \textbf{Material} & \centering \textbf{I$_\textbf{sc}$ $(nA)$} & \centering \textbf{V$_\textbf{oc}$ $(V)$} & \textbf{Ref.}\\
    \hline  
    \centering ZnO & \centering $28.8$ & \centering $1.26$ & \text{{\color{blue} \cite{xu2010self}}} \\
    \hline
    \centering PZT & \centering $-$ & \centering up to $8$ & \text{{\color{blue} \cite{jin2018high}}} \\
    \hline        
        
    \centering BT Arrays & \centering $~0.316$ & \centering $~0.085$ & \text{{\color{blue} \cite{koka2014vertically}}} \\
    \hline
    \centering PVDF & \centering $6-40$ & \centering $0.5-1.5$ & \text{{\color{blue} \cite{persano2013high}}} \\
    \hline
    \end{tabular}
    \end{center}
\end{table}

As the performance of the piezoelectric material is tremendously affected by the material selection, researchers work on a wide variety of piezoelectric materials, both natural and artificial. 
The first material used was quartz in 1880 by Curie Brothers \text{{\color{blue} \cite{inbook}}}; after that, works on piezoelectric materials increased gradually. 
Different types of piezoelectric materials exist, such as inorganic materials, organic materials, and composites. 
PZT, barium titanate $BaTiO_{3}$ (BT); polyvinylidene fluoride polymer (PVDF) \text{{\color{blue} \cite{Kawai_1969}}}, Poly (l-lactic acid) (PLLA) \text{{\color{blue} \cite{zhu2017electrospinning}}}; PZT/PVDF, PVDF/BT \text{{\color{blue} \cite{narita2018review}}} are examples of inorganic, organic and composites piezoelectric materials respectively. 
Short circuit current $(I_{sc})$ and open circuit voltage $(V_{oc})$ of some of the commonly used piezoelectric nanogenerators are given in Table \ref{table:mert_pie}.    
A more detailed investigation of piezoelectric materials is presented in \text{{\color{blue} \cite{SEZER2021105567, narita2018review, ali2019piezoelectric}}}. 
Due to their ability to generate surface charge upon applied stress, the human body is a convenient input source for piezoelectric nanogenerators. 
Piezoelectric nanogenerators can use sources such as heartbeat, lung motion, the flow of blood, and muscular contraction to generate necessary electrical power for biomedical devices, such as cardiac pacemakers \text{{\color{blue} \cite{parvez2018recent}}}, brain stimulators \text{{\color{blue} \cite{zheng2017recent}}}, and respirators \text{{\color{blue} \cite{liu2021piezoelectric}}} to operate. 
More details on piezoelectric energy harvesting systems used for biomedical systems can be found in \text{{\color{blue} \cite{panda2022piezoelectric}}}.
A flexible and miniaturized Piezoelectric Ultrasound Energy Harvesting (PUEH) device is developed in \text{{\color{blue} \cite{zhang2022piezoelectric}}} using samarium (Sm)-doped  Pb(Mg$_{1/3}$Nb$_{2/3}$)O$_{3}$-PbTiO$_{3}$ (Sm-PMN-PT) which can produce an instantaneous output power density of up to 1.1 W/cm$^2$. 
The output power (P) can be expressed as:
\begin{align}
    P   \propto   \frac {\varepsilon_{33}} {c_{33}}   \frac{1}{(k_{33}+\frac{1}{k_{33}})^2} ,
\end{align}
where $\varepsilon_{33}$ is the piezoelectric material’s dielectric coefficient, c$_{33}$ is the effective elastic coefficient, and k$_{33}$ is electromechanical coupling coefficient. 

In general, PENGs possess a greater energy density than TENGs. 
This means that a PENG can generate more energy per unit of volume or surface area. 
PENGs demonstrate efficacy in environments characterized by persistent mechanical tension or vibration. 
Given that numerous IoNT applications may function in such conditions, PENG-based IoHNTs can provide a consistent and dependable power supply wirelessly. 
PENGs, as opposed to TENGs, do not require frictional contact. 
This may reduce the susceptibility of PENGs to wear and strain during extended use. 
Nevertheless, the selection between PENG and TENG ought to be determined by the particular demands/positions of the IoNTs, considering environmental factors, the characteristics of the accessible mechanical energy, and the power specifications of IoNTs. 
A hybrid strategy that combines PENG and TENG technologies may be a more effective solution in certain circumstances.

\begin{figure}[t!]
    \centering
	\hspace{0.1mm}
        \begin{subfigure}[b]{0.19\textwidth}
		{\includegraphics[width=\textwidth]{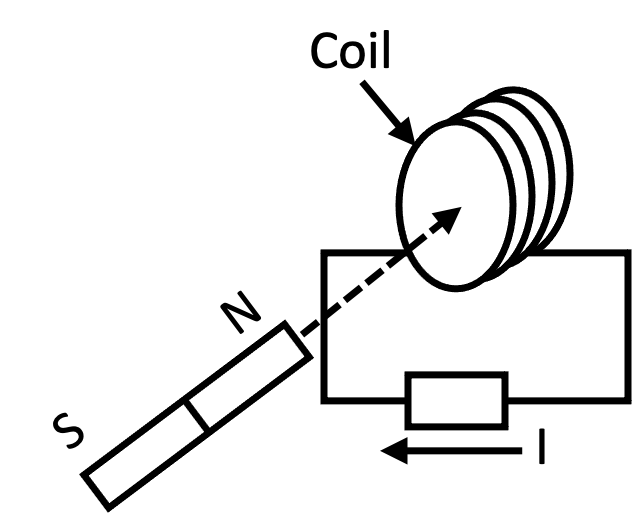}}
		\caption{}
	\end{subfigure}
	\hspace{5mm}
		\begin{subfigure}[b]{0.25\textwidth}
		{\includegraphics[width=\textwidth]{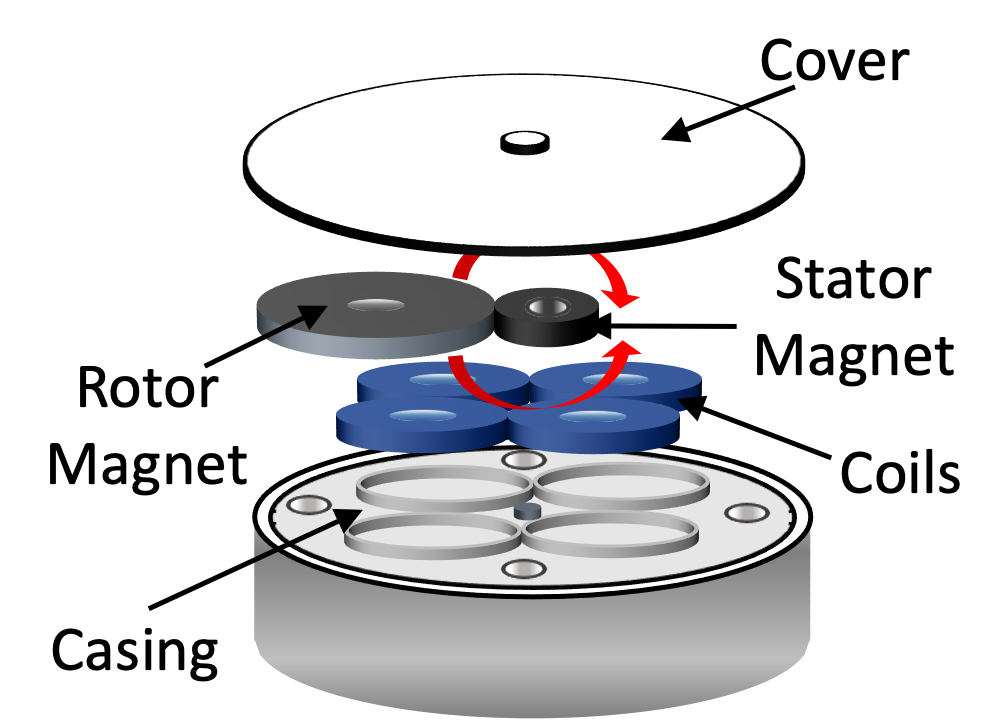}}
		\caption{}
	\end{subfigure}
	\hspace{0.1mm}
	\caption{Electromagnetic Induction Harvester. (a) Working mechanism of electromagnetic induction. (b) A rotational electromagnetic energy harvester (REH) (adapted from \text{{\color{blue} \cite{liu2018non}}}).}
    \label{generalemf}
\end{figure}

\paragraph{Electromagnetic Induction}
Electromagnetic induction phenomena can convert kinetic (mechanical) energy into electrical energy, as shown in Fig. \ref{generalemf}a. 
This phenomenon states that if a magnet moves towards the conducting coil connected with a circuit, the magnetic flux increases. 
Then, an electric current is induced inside the coil due to the change in magnetic field by a magnet \text{{\color{blue} \cite{du2014review}}}. 
Thus, the conversion of mechanical energy to electrical energy is realized, and electromagnetic induction voltage occurs \text{{\color{blue} \cite{mishu2020prospective}}}:
\begin{align}
    \varepsilon_{emg}=-N\frac{d\varphi_{B}}{dt} ,
\end{align}
where $\varepsilon_{emg}$ is the electromotive force (EMF), $\varphi_{B}$ is the magnetic flux, and $N$ is the number of coil turns. 
Magnetic flux is a measurement of the total magnetic field that passes through a given area \text{{\color{blue} \cite{torres2017electromagnetic}}}. 
Electromagnetic Induction EH (EIH) is advantageous compared to other EH methods since it needs only magnets and coils to generate electrical energy, and it is cheap in terms of cost \text{{\color{blue} \cite{bouendeu2010low}}}. 
EIHs have been used frequently in recent years to obtain energy from human movements. 
For instance, a rotational electromagnetic energy harvester (REH), which consists of a disk-shaped rotor magnet, a stator magnet, and four wound coils, is suggested in \text{{\color{blue} \cite{liu2018non, lin2017rotational}}}. 
When a person starts to walk or run with REH on the person's ankle or wrist, the rotor also starts to rotate around the stator, as shown in Fig. \ref{generalemf}b. 
This rotation induces the coils to generate electrical energy based on the magnetic induction phenomena. 
Moreover, other applications of EIHs are integrated with human body parts to convert kinetic energy into electrical energy while running, walking, jogging, or exercising. 
For instance, EIHs can attach to the upper arm \text{{\color{blue} \cite{li2020ultra}}}, ankle \text{{\color{blue} \cite{anjum2018broadband}}} and lower leg \text{{\color{blue} \cite{wang2017magnetic}}}. 
When a person travels, reads a book, works on a desk, or watches TV, movements are less frequent than during exercise \text{{\color{blue} \cite{salauddin2017design}}}. 
Therefore, harvesting energy from motions that are in a low-frequency (6 Hz) range, such as shaking the finger \text{{\color{blue} \cite{kim2019electromagnetic}}} or hand \text{{\color{blue} \cite{salauddin2017design}}} becomes crucial. 
Some researchers use the Halbach magnet array consists of arranged magnets to overcome this problem, described in \text{{\color{blue} \cite{salauddin2017design, kim2019electromagnetic}}}. 
A wearable electromagnetic energy harvester with a conductive textile coil and magnets is demonstrated in \text{{\color{blue} \cite{lee2019wearable}}}.     
An electromagnetic EH to harvest from the bending of soles is proposed in \text{{\color{blue} \cite{wang2022novel}}}. 
During a slow walk (4 km/h), average output power is 9.8 mW with the maximum power of 0.26 W. 

Energy from diverse sources, including fluid flow, ambient vibrations, and rotational motion, can be converted into electrical energy by using EIHs. 
The considerable adaptability of IoNT devices is crucial, as they can be used in various environments utilizing distinct forms of energy with minimal modification. 
EIHs are typically more resilient and require less upkeep than alternative energy harvesting technologies such as TENG and PENG. 
This dependability is of the utmost importance for IoNT devices, which may be installed in remote or inaccessible areas, and information can be gathered remotely and transmitted wirelessly. 

\paragraph{Electrostatic Induction}
Electrostatic induction is the redistribution of the charges when two charged surfaces contact each other to convert mechanical (vibrations) energy into electrical energy. 
Electrostatic energy harvesters (EEH) consist of two conductive plates separated by air or dielectric material as capacitors \text{{\color{blue} \cite{khan2016state}}}. 
When the relative motion occurs between these two plates, electrical energy is generated by the effect of the changing capacitance. 
However, EEHs need an external voltage source and well-designed conversion and extraction circuits to generate electricity sufficiently \text{{\color{blue} \cite{khan2016state}}}. 
Since the output of EEHs is AC, the design of external conversion circuits is crucial in storing the electrical energy effectively. 
There are some studies to overcome this problem in \text{{\color{blue} \cite{kempitiya2011analysis, phan2015low}}}.

\begin{figure}[t!]
	\centering
	{\includegraphics[width=0.40\textwidth]{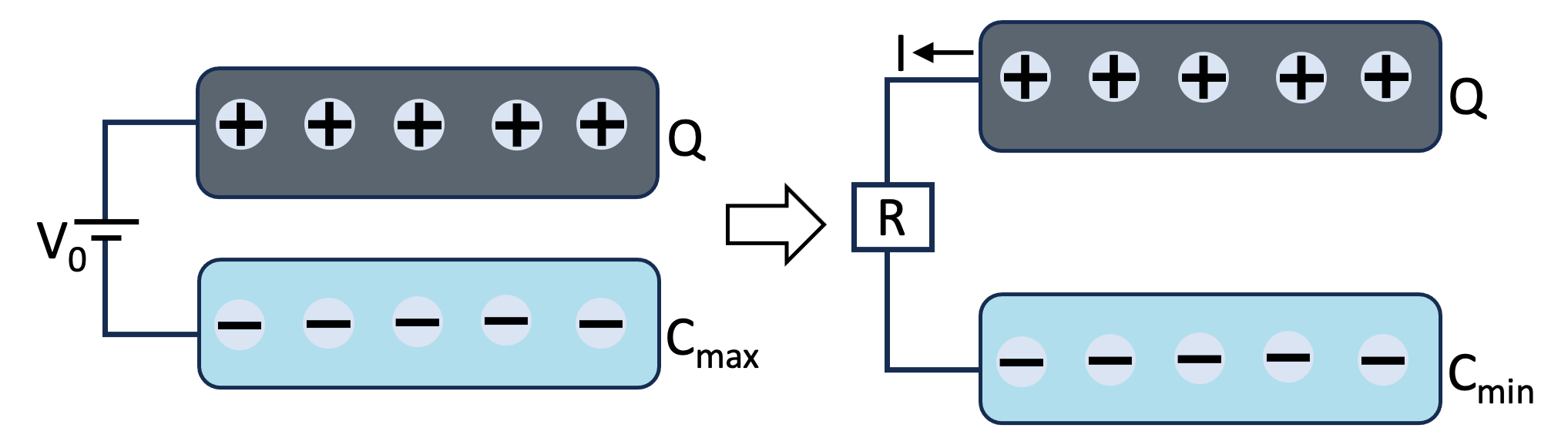}}
	\caption{Charging the conductive plates by using an external source (adapted from \text{{\color{blue} \cite{khan2016state}}}).}
    \label{workingEEH}
\end{figure}

\begin{figure}[t!]
    \centering
    \includegraphics[width=0.3\textwidth]{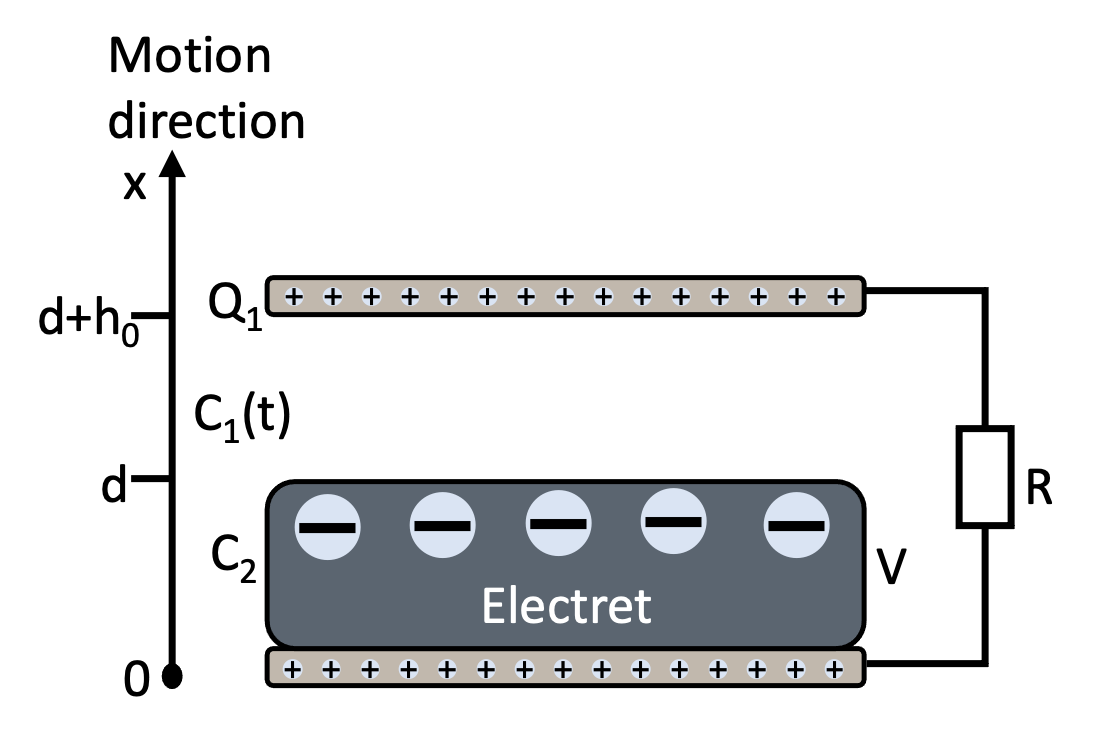}
    \caption{Working principle of the electret-based EEH (adapted from \text{{\color{blue} \cite{zhang2016electrostatic}}}).}
    \label{elecret}
\end{figure}

EEHs consist of two groups: electret-free and electret-based EEHs \text{{\color{blue} \cite{khan2016state}}}. 
An electret is a dielectric material that maintains its electric polarization after exposure to a strong electric field \text{{\color{blue} \cite{boland2003micro}}}, meaning that one side of the electret is quasi-permanently positive, and the other has negative charges. 
In electret-free EEHs, the air is between the conductive plates rather than dielectric material. 
There are two ways to generate electrical energy in electret-free EEHs. 
One way to charge the conductive plates using an external source is to generate a voltage across the plates when relative motion occurs \text{{\color{blue} \cite{khan2016state}}}, as shown in Fig. \ref{workingEEH}. 
The total amount of energy (E) at each cycle is expressed as follows \text{{\color{blue} \cite{khan2016state}}}:
\begin{align}
    E=\frac{1}{2}Q_{0}^2\left(\frac{1}{C_{min}}-\frac{1
    }{C_{max}}\right) ,
\end{align}
where the $Q_{0}$ is the stored charge, $C_{min}$ and $C_{max}$ refer to the minimum and maximum capacitance, respectively, between the plates during relative motion. 
The voltage is generated across the plates when the plates move relatively depending on the energy source (battery or supercapacitor) \text{{\color{blue} \cite{khan2016state}}}. 
The total amount of energy at each cycle is expressed as follows \text{{\color{blue} \cite{khan2016state}}}:
\begin{align}
    E=V_{0}^2(C_{max}-C_{min}) ,
\end{align}
where the $V_{0}$ is the initial charging voltage. Although the working mechanism of electret-based EEHs is quite similar to electret-free EEHs, the main difference between them is the electret layers placed on one or both of the conductive plates, as shown in Fig. \ref{elecret}. 
After putting electrets on the plates, they are charged to convert mechanical vibrations into electrical energy. 
The equation obtained by the electrostatic system is expressed as follows \text{{\color{blue} \cite{zhang2016electrostatic}}}:
\begin{align}
   R\frac{dQ_{1}}{dt}=V-Q_{1}\left(\frac{1}{C_{1}(t)}+\frac{1}{C_{2}}\right) ,
\end{align}
where the $R$ is the external load resistance, $Q_{1}$ is the induced charge on the upper electrode, $V$ is the surface potential of the pre-charged electret, and $C_{1}(t)$ and $C_{2}$ is the capacitance of the air gap and electret dielectric material, respectively.

\begin{figure}[t!]
	\centering
	\begin{subfigure}[b]{0.24\textwidth}
        \centering
		{\includegraphics[width=\textwidth]{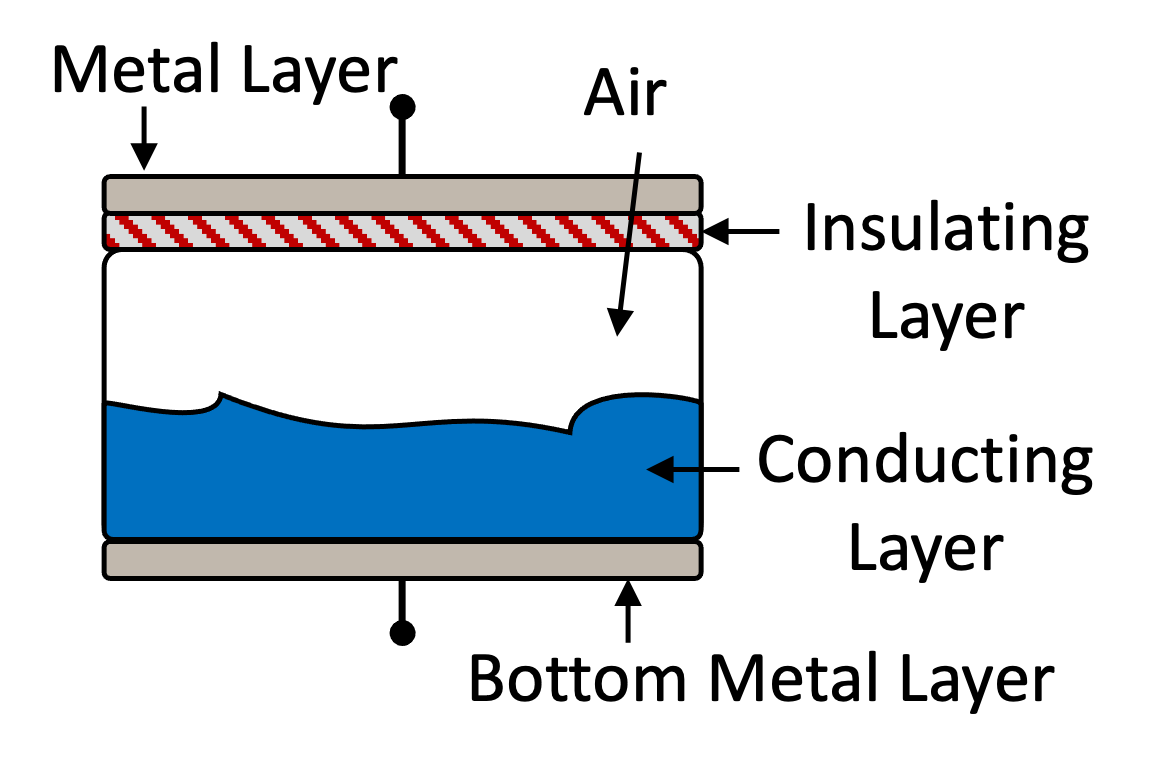}}
		\caption{}
	\end{subfigure}
 \hspace{0.1mm}
	\begin{subfigure}[b]{0.21\textwidth}
        \centering
		{\includegraphics[width=\textwidth]{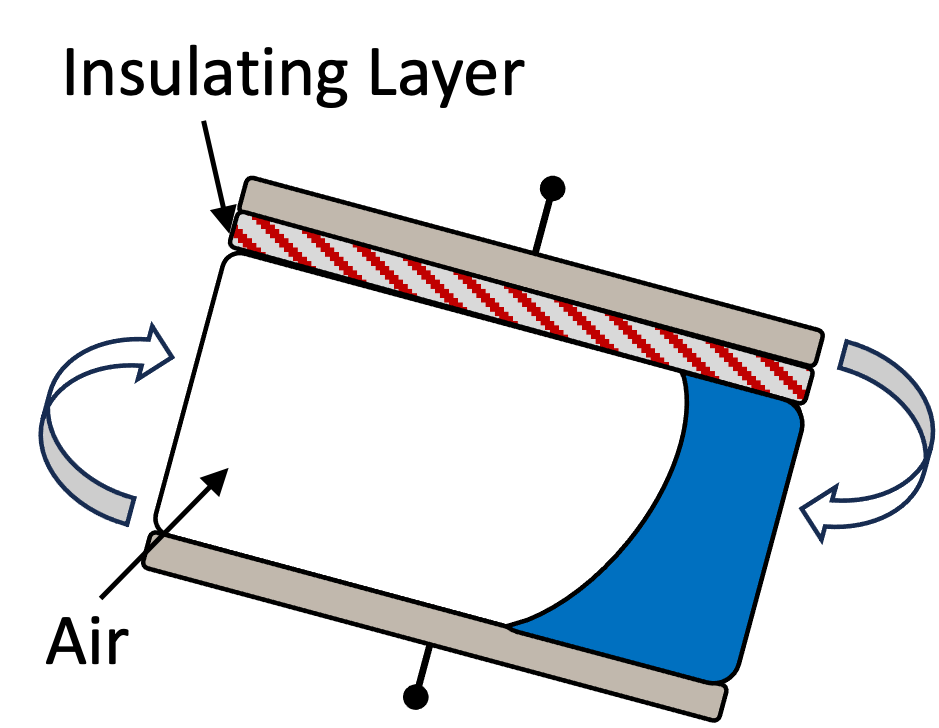}}
		\caption{}
	\end{subfigure}
	\hspace{0.1mm}
	\caption{Example of an EEH that is sensitive to the human motions (adapted from \text{{\color{blue} \cite{choi2011liquid}}}). (a) Structure of the EEH. (b) The movement of the liquid in the harvester causes the capacitance change.}
    \label{expEEH}
    \end{figure}

Most environmental energy sources and human body movements are in a low-frequency range of less than 100 Hz.
Hence, there are some studies that enhance the performance of EEHs at low frequencies. 
An EEH that includes a conducting liquid between the plates rather than the conventional EEH structure to enhance the sensitivity of the EEH is proposed in \text{{\color{blue} \cite{choi2011liquid}}}, as shown in the Fig. \ref{expEEH}a. 
When the motion occurs, the movement of the liquid in the harvester causes the capacitance change, which provides obtaining energy as a result, as illustrated in Fig. \ref{expEEH}b. 
EEHs exhibit notable efficiency in environments characterized by low-frequency vibrations. 
Such environments are frequented by numerous IoT and IoNT applications, which makes EEHs an ideal device for efficiently capturing and converting ambient energy to electrical energy and wirelessly power IoNTs. 

\subsubsection{Body Heat}
To harvest energy from the human body's movements, a person must be in motion for a certain amount of time during the day. 
On the other hand, heat is always in the human body and an accessible energy source to harvest energy. 
Usually, the environment temperature is around 20-25${^o}$C, and the human body temperature is around 37${^o}$C \text{{\color{blue} \cite{nozariasbmarz2020thermoelectric}}}. 
This temperature difference between the environment and the human body can be used as the source for energy harvesting \text{{\color{blue} \cite{nozariasbmarz2020thermoelectric}}}.

\begin{figure}[t!]
\centering
\includegraphics[width=0.4\textwidth]{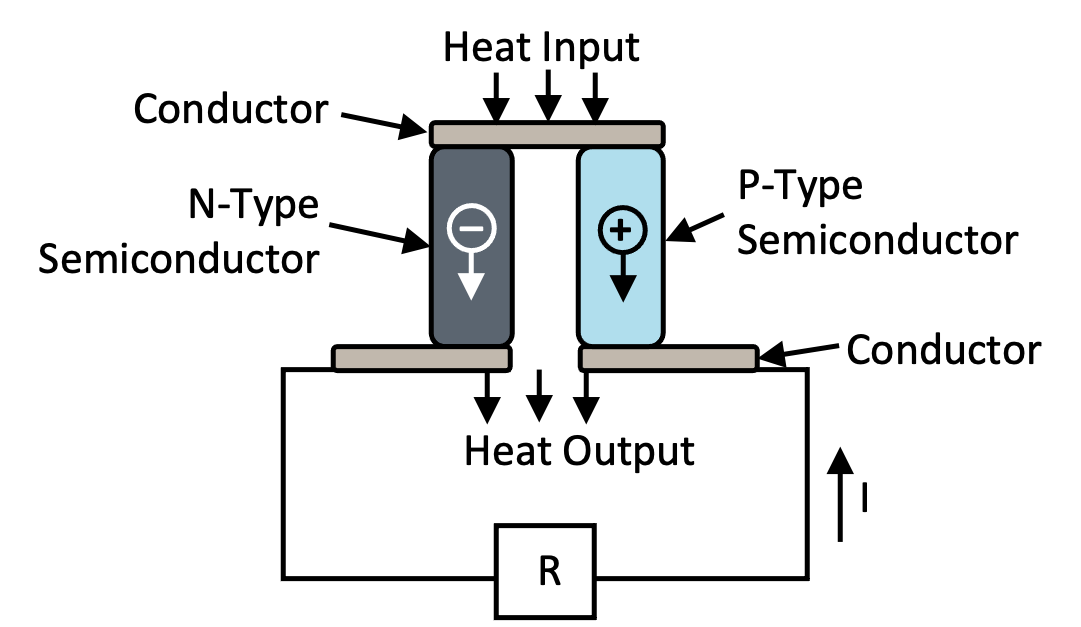}
\caption{Seebeck effect demonstration (adapted from \text{{\color{blue} \cite{thermo_rev}}}). }
\label{fig:seebeck}
\end{figure}

\paragraph{Thermoelectric Effect}
Thermoelectric generators (TEG) generate electricity by direct contact with the human body. 
The generation occurs with the Seebeck effect, in which the electrical potential difference emerges from the temperature difference between the junctions of the TEG, which is demonstrated in Fig. \ref{fig:seebeck}. 
The generated voltage through a TEG can be found as follows:
\begin{equation}
    V = n\alpha\Delta T_{TEG},
\end{equation}
where $n$, $\alpha$, $\Delta T_{TEG}$ corresponds to the number of thermocouples, the Seebeck coefficient of the selected thermoelectric material, and the temperature gradient between the junctions, respectively \text{{\color{blue} \cite{see}}}. 
Since the resistivity of TEG is generally small concerning air and body resistance, the temperature difference between ambient air and the human body is generally higher than that of TEG, $\Delta$$T_{TEG}$ becomes lower than the measured temperature difference between the air and the human body. 
However, environmental factors, such as ambient temperature, wind speed, clothing thermal insulation, and a person's activity, can affect performance significantly. 
For example, it is demonstrated that human activities, such as walking, cycling, and sitting, result in a different amount of electrical power generation, which ranges from 5 to 50 $\mu$W \text{{\color{blue} \cite{thermo}}}. 
Another critical point to consider when harvesting energy through TEG is where to put it. 
Since the head, chest, and wrist have the highest temperatures, they are the most energy-efficient parts of the body to generate electricity through heat.
The average recorded temperature of the different parts of the human body is shown in Table \ref{table:mert1}. 
From Table \ref{table:mert1}, it can be deduced that due to the more significant temperature difference between the skin surface and the environment, harvesting conducted at 17$^o$C is a better choice in terms of efficiency.

\begin{table}[t!]
    \centering
    \caption{Temperatures of human body parts recorded at different ambient temperatures}
    \label{table:mert1}
    \begin{tabular}{  |>{\centering\arraybackslash} p{2.2cm}  |>{\centering\arraybackslash}m{1.9cm}  |>{\centering\arraybackslash} m{1.9cm} |>{\centering\arraybackslash} m{1.9cm}|}
    \hline
       
    \textbf{Body  Part} &  \textbf{17.0$\boldsymbol{^\circ}$C} \text{{\color{blue} \cite{Hee2011ASO}}} &  \textbf{23.5$\boldsymbol{^\circ}$C} \text{{\color{blue} \cite{Zaproudina_2008}}} & \textbf{27.0$\boldsymbol{^\circ}$C} \text{{\color{blue} \cite{Webb2004TemperaturesOS}}}\\
            
    \hline
    Thigh &  28.3   & 30.8 &   33.0 \\
    \hline
    Calf &   29.4  & 31.3   & 31.6 \\
    \hline
    Arm anterior & 30.3 & 31.7 &  33.2\\
    \hline
    Forearm    & 29.5 & 31.5 &  34.0\\
    \hline
    Dorsum of foot & 27.1  & 28.6   & 30.5\\
        
    \hline
        
    \end{tabular}
        
\end{table}

The circuitry arrangement, especially impedance matching, is essential to obtain the maximum power to be delivered on the load. 
The maximum power obtained through a thermopile can be found as follows: 
    
\begin{equation} 
\label{maxpo}
P_{max} = \frac{1}{4}ZW^{2}R_{tp}, 
\end{equation}
where $P_{max}$ is the maximum power on the matched load, $Z$ is the thermoelectric figure of merit, $Z = S^2\sigma/k$, $S$ is the Seebeck coefficient, $\sigma$ and $k$ are electrical and thermal conductivities, respectively, $W$ is the heat flow over the thermopile, which is equal to is equal to $\Delta T R_{tp}$, and $R_{tp}$ is its effective thermal resistance when $P_{max}$ is reached. 
From \textcolor{blue}{(\ref{maxpo})}, it is clear that to maximize the  $P_{max}$, one should select or design a material having a large value of $Z$. 
However, due to $R_{tp}$ being a nonlinear function of $Z$, one must be careful, as attempts increasing $Z$ will cause a drop in $R_{tp}$ based on the Peltier effect after the optimum value of $Z$ is determined based on the maximum electrical power. 
According to \textcolor{blue}{(\ref{maxpo})}, $W$ and $\Delta T$ depend on the thermal resistance of a heat source and a heat sink. 
It is also essential to state that by the thermal impedance matching theory, a wearable thermoelectric energy harvester (TEH) must have high thermal impedance \text{{\color{blue} \cite{matching}}}. 
However, thermal impedance of the TEH and the environment must be matched to obtain the maximum power from a TEH. 

A comprehensive analysis of various material options for TEG is presented in \text{{\color{blue} \cite{mat}}}. 
There are many applications of TEG. 
It can be implemented on a cloth, such as a shirt, as in \text{{\color{blue} \cite{heat}}}, which does not require any further modifications. 
Yet, the implementation methodology is so simple that just glue and sewing would be enough. 
A technique is proposed to cultivate energy from the human body temperature gradient with the help of PEDOT: PSS/ CNT@PDA@PDMS (PCPP) sensor in \text{{\color{blue} \cite{gao2022integrated}}}. 
The harvested energy can be used in various applications, such as health monitoring \text{{\color{blue} \cite{thermo_mon}}}, fault detection \text{{\color{blue} \cite{fault}}}, Internet of Things \text{{\color{blue} \cite{nod, 9174941}}},  and powering health sensors \text{{\color{blue} \cite{thermo_sen}}}.

The human body generates a consistent and easily accessible supply of energy in the form of body heat. 
TEHs have the ability to consistently transform the heat generated by the human body into electrical energy, thereby serving as a dependable power supply for wearable and implantable IoNT devices. 
TEHs are non-invasive and safe to be used in direct contact with the human body. 
Biomedical IoNT devices necessitate this attribute as it is of utmost importance to ensure biocompatibility and minimize potential risks to the user. 
TEHs need minimum to no maintenance, which is especially advantageous for IoHNT-based IoNT systems that are designed for prolonged utilization in medical or health monitoring contexts. 
These characteristics make them well-suited as IoHNT utilized in the fields of health monitoring, fitness tracking, and medical diagnostics.

\paragraph{Pyroelectric Effect}
Like TEGs, pyroelectric generators (PGs) are also used to harvest energy from the temperature difference. 
However, the PG’s work is based on the phenomenon of pyroelectricity. 
According to this phenomenon, temperature fluctuations with time in pyroelectric material is considered rather than the spatial temperature gradient as in the thermoelectric effect \text{{\color{blue} \cite{thakre2019pyroelectric}}}. 
Therefore, the pyroelectric coefficient ($P_{pyro}$) and output current ($I$) can be given as follows \text{{\color{blue} \cite{mishu2020prospective}}}:
\begin{align}
    P_{pyro}&=\frac{d\rho}{dT},\\[6 pt]
    I=\frac{dQ}{dT}&=\mu P_{pyro}A\frac{dT}{dt},
\end{align}
where the $\rho$ is the spontaneous polarization, $T$ is the temperature, $Q$ is the induced charge, $\mu$ is the absorption coefficient, and $A$ is the surface area. 
Pyroelectric materials that exhibit spontaneous polarization even if there is no applied electrical field are a subdomain of the piezoelectric materials \text{{\color{blue} \cite{bowen2014pyroelectric}}}. 
Therefore, PGs generally consist of ferroelectric materials (PZT, PVDF, etc.) that exhibit more significant pyroelectric and piezoelectric properties than non-ferroelectric materials \text{{\color{blue} \cite{korkmaz2021pyroelectric}}}. 
A typical PG comprises a pyroelectric material sandwiched between two metal electrodes. 
When PG is heated, the dipole alignment in the pyroelectric material disturbances occur due to the thermal vibration and, therefore, electrical polarization changes. 
This change leads to the flow of electrons to the electrode surface by separating bound charges. 
When PG is cooled, dipoles return to their spontaneous polarization so that electrical polarization increases and electron flow occurs in the reverse direction \text{{\color{blue} \cite{korkmaz2021pyroelectric}}}. 
The pyroelectric effect depends on the rate of temperature change. 
Since the cooling rate is lower than the heating rate, PG's output current or voltage has a larger amplitude when PG is heated \text{{\color{blue} \cite{sultana2018pyroelectric, bhatia2014high}}}.
Although EH with PGs is studied less than thermoelectric and piezoelectric EH, many applications in the literature with PGs exist. 
For instance, in continuous health monitoring systems, PGs can be a potential energy harvester for thermosensors and breathing sensors to detect human body heat \text{{\color{blue} \cite{zhang2016flexible}}} and monitor the respiratory process \text{{\color{blue} \cite{sultana2018pyroelectric}}}. 

PEHs are efficient in situations characterized by temperature variations, such as those experienced with body heat throughout various activities or environmental circumstances. 
IoNTs are commonly utilized in dynamic, real-world settings characterized by frequent temperature fluctuations, such as when placed on the human body. 
In such scenarios, PEHs can consistently extract energy from these variations, guaranteeing a stable and uninterrupted energy transmission to IoNTs. 
Pyroelectric IoHNTs are devoid of any mechanical components, leading to less friction and prolonged operational duration. 

\subsection{Biofuel Cell Energy Sources}
Instead of the energy directly generated by the human body itself, we can use the body as a place to generate power through biofuel cells (BFC).  
We can categorize the BFCs into two sub-groups: enzymatic fuel cells (EFC) and microbial fuel cells (MFC). 
A comparison between EFC and MFC is tabulated in Table \ref{table:mert2}.
The electrical power generation in BFCs occurs through the oxidation of the substrate material. 
As living organisms constantly have substrates, such as glucose and alcohol, that can be oxidized to generate electricity, the generation process will continue without external interference. 
Oxygen at the cathode oxidizes the anode's oxidant, and water, heat, and electrons emerge as by-products via the electrochemical reaction. 

\begin{table}[t!]
\caption{Differences between EFC and MFC}
\centering
\begin{tabular}{|>{\centering\arraybackslash} m{2.8cm} |>{\centering\arraybackslash} m{3cm} |>{\centering\arraybackslash} m{3.2cm}|}
    \hline
    \textbf{Features} & \textbf{Enzymatic Fuel Cells (EFCs)} & \textbf{Microbial Fuel Cells (MFCs)} \\
    \hline
    Energy Source & Organic compounds (e.g., glucose) & Microorganisms (bacteria, algae) \\
    \hline
    EH Mechanism  & Enzymatic reactions & Microbial metabolism \\ \hline
    Ambience & Requires controlled conditions for optimal function & Can operate in diverse environments, e.g., wastewater \\ \hline
    Controllability	& High due to precise enzyme reactions & Limited as depends on environmental factors \\ \hline
    Substrate Utilization & Limited to specific substrates & Can utilize a wide range of organic compounds\\ \hline  
    Power Output & 12$-$483 $\mu$W/cm$^2$ \bluecite{huang2022transient, zhiani2022ex} & 0.1$-$200 $\mu$W/cm$^2$ \bluecite{koffi2020high}\\ \hline
    Power Conversion Efficiency & $-$ & Maximum 83\% \bluecite{koffi2020high} \\ \hline
    Advantages & Rapid, bio-compatible & Compatible in diverse environment, bio-compatible \\ \hline
    Disadvantages & Limited stability, Substrate specific & Uncertainties in complex microbial interactions \\ \hline
    IoNT application areas & Healthcare, Industry & Environment, Industry \\ \hline
\end{tabular}
\label{table:mert2}
\end{table}

\subsubsection{Enzymatic Fuel Cells (EFCs)}
\begin{figure}[t!]
    \centering
    \includegraphics[width=8cm]{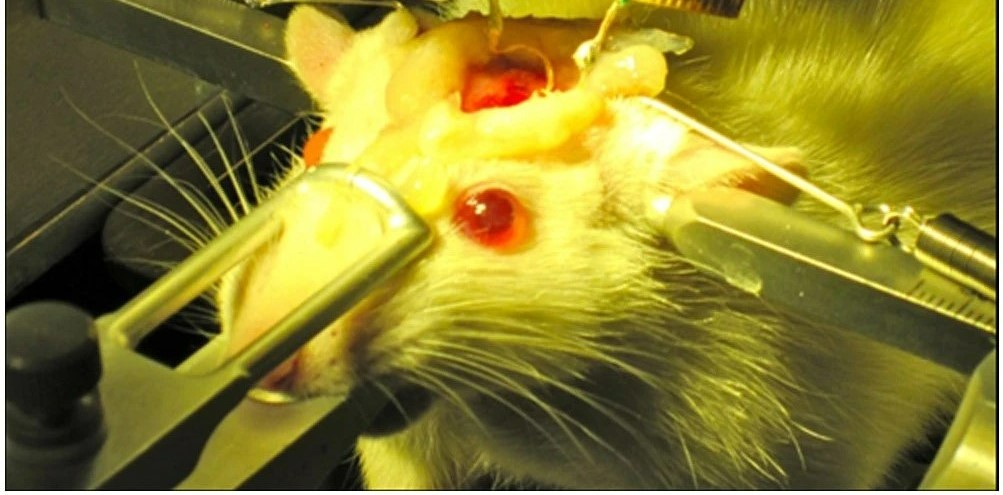}
    \caption{EFC implementation in a rat \text{{\color{blue} \cite{rat-fig}}}.}
    \label{fig: rat}
\end{figure}

Works on the EFCs started in the 1960s after the realization of the possibility of using electrochemical enzymes to catalyze the fuels at the anode. 
After approximately 40 years, the first implantable Glucose BFC (GBFC) is produced in \text{{\color{blue} \cite{heller}}}, which is implanted in a grape as a profound step to integrating EFCs with living beings. 
Then, one by one, researchers injected the GBFC into living organisms, such as rat \text{{\color{blue} \cite{rat}}}, snail \text{{\color{blue} \cite{snail}}}, and lobster \text{{\color{blue} \cite{bfc-lob}}}. 
An example of EFC implementation in living organisms can be seen in Fig. \ref{fig: rat}.
An enzymatic biofuel cell using single reduced graphene oxide (rGO)-glucose oxidase bioanode (GOx/rGO) and a single rGO-laccase (Lac/rGO) biocathode, which provides a maximum power density of 4.0 nW/cm$^{2}$ and an open-circuit voltage of 0.04 V is demonstrated in \text{{\color{blue} \cite{kabir2022energy}}}.

Transient glucose enzymatic biofuel cells (TEBFCs) with laser-induced graphene(LIG)/gold nanoparticles (Au NPs) composite electrodes are proposed in \text{{\color{blue} \cite{huang2022transient}}}, which shows an output power density of 483.1 $\mu$W/cm$^{2}$. 
An ex vivo EH technique, which is the development of abiotically catalyzed Glucose Fuel Cells (AGFCs), is demonstrated in \text{{\color{blue} \cite{zhiani2022ex}}}, that offers high performance, produces a power density of ~12.5 $\mu$W/cm$^{2}$. 
The established half-life time of the AGFC device is 8 hours \text{{\color{blue} \cite{zhiani2022ex}}}. 
Along with glucose, hydrogen, lactate, and alcohol can also be used to produce electrical power. 
Energy harvesting from the human sweat presented in \text{{\color{blue} \cite{perspiration}}}.
Using lactate, abundantly found in human sweat, as the biofuel, the electrical power is generated via epidermal BFCs based on temporary transfer tattoos (tBFCs). 
The selection of anode and cathode materials can cause the generated output power to differ. 
Some of the materials and the produced electrical power are shown in Table \ref{table:mert3}. 
A large number of BFC applications exist for powering electronic devices. 
For example, a pacemaker's battery is replaced with external BFCs in \text{{\color{blue} \cite{bfc-lob}}}. 
Similarly, EFCs can power wireless electronic devices  \text{{\color{blue} \cite{falk2014self, hanashi2011bioradiotransmitter}}}. 
A combination of glucose-based BFC and brain stimulator implementation in a pigeon is presented in \text{{\color{blue} \cite{LEE2021112746}}}.
\begin{table}[t!]
\centering
\caption{Performance Metrics of EFCs}
    \label{table:mert3}
        \begin{tabular}{|>{\centering\arraybackslash}m{1.3cm} |>{\centering\arraybackslash}p{1.2cm} |>{\centering\arraybackslash}m{1.3cm} |>{\centering\arraybackslash}m{1.3cm} |>{\centering\arraybackslash}m{1.7cm} |>{\centering\arraybackslash}m{0.7cm}|}
        \hline
        \textbf{Anode} & \textbf{Cathode} &  \textbf{Current Density $(\mu A/cm^{2})$}  & \textbf{Power Density $(\mu W/cm^{2})$} & \textbf{Open Circuit Voltage} (V)& \textbf{Ref.}  \\
        \hline
        Methanol   &  Pt  &  2600 & 680 & 0.8 & \text{{\color{blue} \cite{palmore1998methanol}}} \\
        \hline
        Ethanol   &  Pt  &  3  & 1160 & 0.62& \text{{\color{blue} \cite{akers2005development}}} \\
        \hline
        Fructose   &  Laccase &  2800 & 850 & 0.79& \text{{\color{blue} \cite{kamitaka2007fructose}}} \\
        \hline
        Glucose   &  Laccase  &  4470   & 1540 & 0.76 &\text{{\color{blue} \cite{reuillard2013high}}} \\
        \hline
        \end{tabular}
\end{table}
Numerical modeling of EFC is still under research \bluecite{osman2013mathematical}; an overview of the mathematical model is given hereunder.  
The rate of electrochemical reactions at the anode and cathode can be expressed by Butler-Volmer equation, given by:
\begin{align}
    r_{a} = A_{a} k_{a} C_{a_{red}}^{\alpha_{cf}} C_{a_{ox}}^{(1-{\alpha_{cf}})}
    \times \left[\exp \left(\frac{(2(1-\alpha_{cf}) F \eta_{a}}{RT} \right) - \exp \left(\frac{(-2\alpha_{cf}) F \eta_{a}}{RT} \right) \right],
    \label{anode}
\end{align}
\begin{align}
    r_{c} = A_{c} k_{c} C_{c_{red}}^{\alpha_{cf}} C_{c_{ox}}^{(1-{\alpha_{cf}})}
    \times \left[\exp \left(\frac{-\alpha_{f} F \eta_{ca}}{RT} \right) - \exp \left(\frac{(1-\alpha_{f}) F \eta_{ca}}{RT} \right) \right],
    \label{cathode}
\end{align}
where $r_{a}$ \& $r_{c}$ are the rate at anode and cathode, respectively, $A_{a}$ \& $A_{c}$ are the anode and cathode specific surface area, respectively, $k_{a}$ \& $k_{c}$ are rate constants at anode and cathode, respectively. $C_{a_{red}}$ and $C_{a_{ox}}$ are concentrations of reduced and oxidized forms of the mediator at the anode, respectively, whereas concentrations of reduced and oxidized forms of the mediator at the cathode are $C_{c_{red}}$ and $C_{c_{ox}}$, respectively. $\alpha_{cf}$ is the charge transfer coefficient, $\alpha_f$ is the transfer coefficient, $F$ is the Faraday's constant, $\eta_{ca}$ and $\eta_a$ are the cathode and anode overpotential, $R$ is the ideal gas constant, and $T$ is the temperature. Overpotential is the difference between the actual potential at which an electrochemical reaction occurs on an electrode surface and the theoretical equilibrium potential of that reaction.
The equilibrium potential of the redox reaction is given by:
\begin{equation}
    E = E_0 -\frac{RT}{n_e F} log \left(\frac{C_{red}}{C_{ox}} \right) - 0.03 n_H (pH-7),
\end{equation}
where $E$ is the standard electrode potential for the reaction, $E_0$ is the standard electrode potential for a reference half-cell, $n_e$ \& $n_H$ are the electrodes and protons transferred, respectively, $C_{red}$ \& $C_{ox}$ are the reduced and oxidized chemical species, respectively, and $pH$ is the acidity of the solution \bluecite{osman2013mathematical}. 

EFCs can play a crucial role in delivering a durable and effective energy source to nanodevices integrated into the IoNT ecosystem. 
EFCs employ enzymes as catalysts to transform biofuels, such as glucose, into electrical energy. 
Due to their great biocompatibility, these materials are well-suited for use in biomedical applications, particularly for implanted or wearable IoNT devices that function in close proximity to or inside biological systems. 
EFCs are efficient in water-based surroundings, a crucial requirement for biomedical IoNT devices that may come into contact with body fluids and are able to directly use chemical energy from these fluids.  
Due to the consistent presence of biofuels in the human body, EFC based IoHNTs have the potential to offer a continuous and essential energy supply for IoNT devices that require autonomous operation for long duration.

\subsubsection{Microbial Fuel Cells (MFC)}
The opportunities brought up by MFC capture the researcher due to the ability to utilize waste produced by living organisms by converting them to electrical power through microbially catalyzed electrochemical processes. 
The usage of microorganisms to generate electricity was first proposed in \text{{\color{blue} \cite{Potter}}}. 
Substrate selection is one of the main factors affecting the generated electrical power output. 
Numerous choices of substrates exist, varying from simple compounds to high-complexity mixtures to be oxidized through the activities of microorganisms \text{{\color{blue} \cite{mfc}}}. 
Although oxygen is the first option for the selection of the oxidant due to its availability and ease of reach, previous research shows that metallic materials can be selected as oxidants as alternatives. 
For example, U, Cd, Cr, and Cu are used in \text{{\color{blue} \cite{u}}}, \text{{\color{blue} \cite{cd}}}, \text{{\color{blue} \cite{cr}}}, and \text{{\color{blue} \cite{cu}}}, respectively. 
Due to their selectivity and biocompatibility, enzymes do not need a membrane. 
Although there exist membrane-free MFCs \text{{\color{blue} \cite{membrane}}}, due to the oxygen diffusion problem, it is not an ideal solution. 
Numerical modeling of MFC is not a well-developed area \bluecite{osman2013mathematical}; still, a conceptual mathematical model that characterizes the working principle of MFC is given. 

The rate of substrate consumption, which is the substrate conversion with an oxidized mediator can be expressed by Double Monod limitation and given by 
\begin{align}
    r = r_{max} \Phi_b \frac{C_a}{K_a + C_a} \frac{C_d}{K_d + C_d}
\end{align} 
where $r$ is the rate of substrate consumption, $r_{max}$ is the maximum rate constant for substrate consumption, $\Phi_b$ is the volumetric fraction of active microbial biomass, electron acceptor concentration is expressed by $C_a$, $K_a$ is the Monod half-saturation coefficient for electron acceptor, electron donor concentration is shown by $C_{d}$, and $K_{d}$ is the Monod half-saturation coefficient for electron donor. The mass balance equation is given by
\begin{align}
    \frac{dC_d}{dt} = \frac{\Phi}{V_{an}} (C_{d0} - C_d) + r_B + \frac{1}{V_{an}} \int\limits_{V_{bf}}r_{bf} \, dV + \frac{1}{V_{an}} \int\limits_{A_{bf}} r_s \, dA
\end{align}  
where $\Phi$ is the volumetric fraction of microbial biomass, $V_{an}$ is anode chamber volume, $C_{d0}$ is initial substrate concentration, $C_{d}$ is substrate concentration, net reaction rate in bulk is given by $r_B$, $V_{bf}$ is volume of biofilm, $r_{bf}$ is net reaction rate in biofilm, $A_{bf}$ is surface area of of biofilm, and $r_s$ is electrochemical rate of solute. It can be solved considering suitable initial and boundary conditions. Current density that depends upon anode overpotential, the potential difference between the thermodynamically occurring potential and the potential at which it actually does, can be expressed by Bulter-Volmer equation, given by
\begin{align}
        J = J_0 \left[ \exp \left( \frac{\alpha n F}{R T} \eta_{an} \right) - \exp \left( -\frac{(1-\alpha) n F}{R T} \eta_{an} \right) \right]
\end{align}
where $J$ is current density, $J_0$ is exchange current density, $\alpha$ is charge transfer coefficient of anodic reaction, $n$ is the number of electrons transferred, $F$ is Faraday's constant, $R$ is ideal gas constant, $T$ is temperature, and $\eta_{an}$ is anode overpotential. Lesser overpotential indicates more efficient MFC.

\begin{figure}[t!]
        \centering
        \includegraphics[width=9cm]{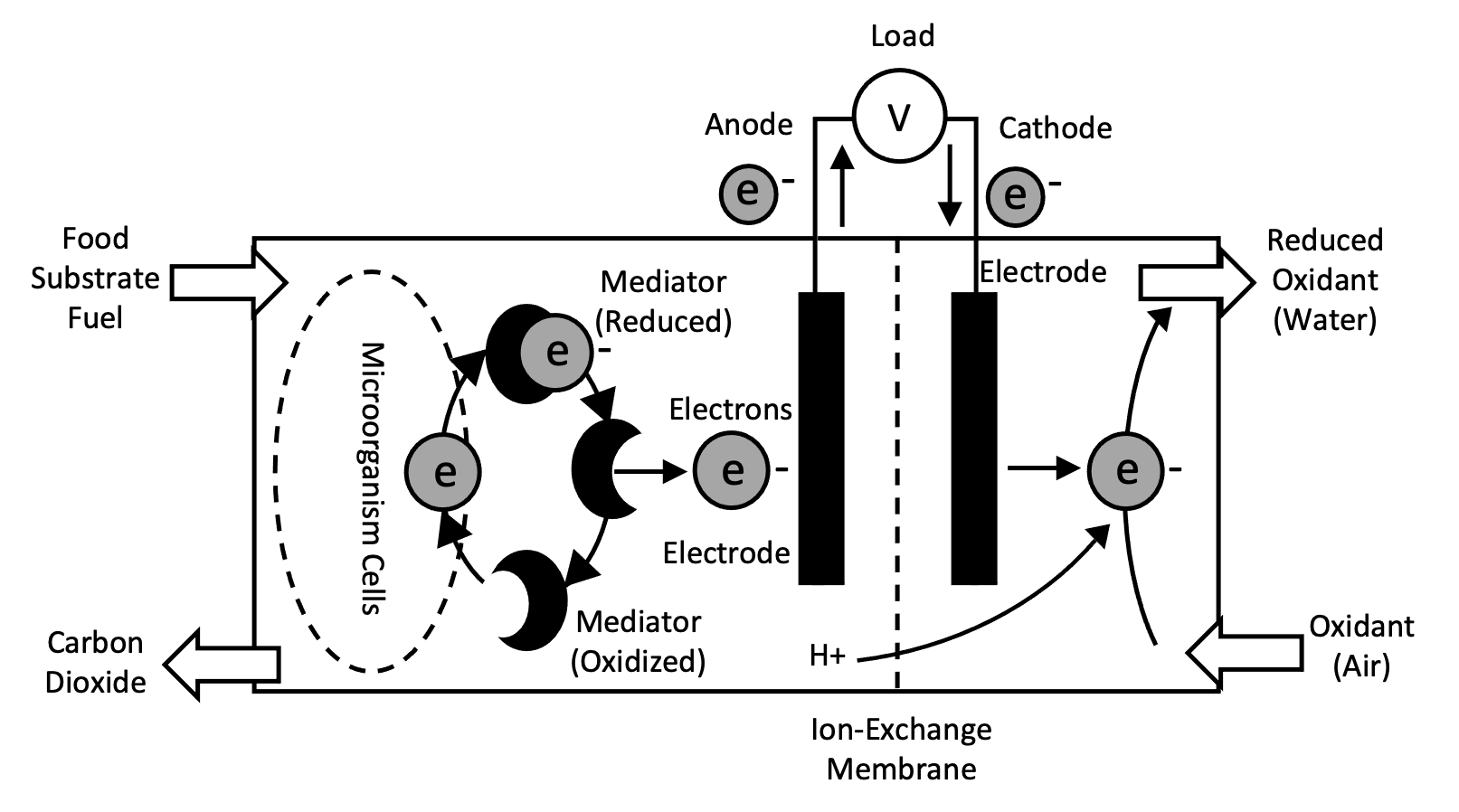}
        \caption{An example model for MFC (adapted from \text{{\color{blue} \cite{wilkinson2000gastrobots}}}).}
        \label{fig: mfc-mod}
\end{figure}

Due to the considerable volumetric size requirement, MFCs are not as practical as EFCs for in vivo applications, yet this is still an open research topic for researchers. 
However, large human intestines can be a suitable place for EH via MFC, as proposed in \text{{\color{blue} \cite{DONG2013916}}}. 
Hence, it is possible to power IoNTs through MFCs in proper situations. 
Another exciting application and way of using MFC in living organisms is EH via MFCs using urine as substrate, as demonstrated in \text{{\color{blue} \cite{Ieropoulos2012UrineUB}}}.
A basic model of how MFC operates is shown in Fig. \ref{fig: mfc-mod}.
Furthermore, an investigation on whether harvested electrical energy from urine can be used to power wearable electronic devices is presented in \text{{\color{blue} \cite{5f97b36643534abe8e0945d418b63c79}}}, where an MFC system is implemented within a pair of socks. 
Sweat is a body fluid whose content is rich in terms of biomarkers. 
It contains ions (Na$^{+}$, Ca$^{+2}$, K$^{+}$, etc.), metabolites (glucose, lactate, ethanol, etc.), hormones (cortisol, etc.), small proteins, and peptides \text{{\color{blue} \cite{sonner2015microfluidics}}}. 
During the exercise, the lactate concentration in the sweat increases, so sweat can be a potential energy source in EH applications due to the lactate's ionic strength.  
For wearability, a textile-based biofuel cell with electrodes implanted into the fabrics can power the LED and electronic watch when a person wears them as a headband and wristband, respectively \text{{\color{blue} \cite{jia2014wearable}}}.

MFCs have the ability to produce electrical energy from organic waste substances, including bodily fluids like sweat, saliva, or urine.  
The human body harbors a wide variety of microbial flora. 
MFCs have the ability to utilize naturally existing bacteria to produce electricity, establishing a mutually beneficial connection between the technology and the human host, which may be used to wirelessly power IoNTs.  
To fully harness the promise of MFCs in powering IoNTs for biomedical applications, it is crucial to tackle challenges related to long-term stability, efficiency, and total biocompatibility.

\subsection{Other Energy Harvesting Types}
A group of researchers are working on MXenes (a two-dimensional transition metal carbide/nitride) for harvesting energy from various ambient energy sources, like thermoelectric energy \text{{\color{blue} \cite{lu2020high, khazaei2013novel}}}, solar energy \text{{\color{blue} \cite{kang2017mxene, yang2019surface}}}, piezoelectric energy \text{{\color{blue} \cite{tan2019large}}}, triboelectric energy \text{{\color{blue} \cite{fan2023mxene}}}, electrokinetic energy \text{{\color{blue} \cite{yang2020ultrathin}}}, salinity gradient energy \text{{\color{blue} \cite{xiao2019ion}}}, mechanical energy \text{{\color{blue} \cite{yang2022fully}}} and ultrasound energy \text{{\color{blue} \cite{lee2020ultrasound}}}. 
However, the research in this domain is still in its infancy and requires significant research in energy harvester development, MXenes synthesis, and energy sources.  

A Body Integrated Self-powered System (BISS) for the energy scavenging purpose from body motions is proposed in \text{{\color{blue} \cite{shi2019body}}}. 
This device can be used for both in vivo and in vitro applications. 
During movement, like walking, jumping, and running, the human body generates electricity by triboelectrification. 
The study shows the feasibility and compatibility with the human body and proved to be suitable for powering wearable and implantable devices. 
This BISS is not only just an EH device but can also be used as a sensor for motion sensing, human activity analysis, and gait analysis \text{{\color{blue} \cite{shi2019body}}}.
An electrostatic MEMS-based EH to provide energy to the battery of leadless pacemakers using vibrations produced by the movement of the heartbeat is elaborated in \text{{\color{blue} \cite{ambia2022microfabrication}}}. 
It can produce an average power of 7$\mu$W. 
In Table \ref{tab:general}, results obtained from different energy sources are tabulated and elaborated. 
These technologies can have a crucial function in wirelessly powering the devices that are integrated into the IoNT ecosystem, especially in the healthcare and industrial sectors. 
IoNTs, such as sensors for health monitoring or diagnostic tools at the nanoscale, frequently require continuous and self-sustained operation. 

\begin{landscape}
\begin{table*}[t!]\centering
  \caption{Comparison among different IoHNT techniques.}
    \label{tab:general}
\scriptsize{
 \setlength{\tabcolsep}{5pt}
 \addtolength{\tabcolsep}{-1pt} 
 \begin{tabular}{|>{\centering\arraybackslash}m{1.8cm} | >{\centering\arraybackslash}m{1.8cm}| >{\centering\arraybackslash}m{2cm}| >{\centering\arraybackslash}m{1.7cm} |>{\centering\arraybackslash}m{2cm}| >{\centering\arraybackslash}m{2cm}| >{\centering\arraybackslash}m{2.7cm}| >{\centering\arraybackslash}m{1 cm}|}
 \hline

\textbf{Source} & \textbf{EH Method} & \textbf{Features} & \textbf{Power Density}  & \textbf{Advantages} & \textbf{Disadvantages} & \textbf{IoNT application areas} & \textbf{Ref.}\\ \hline
\vspace{-1.0cm}

\textit{Sunlight} & Photovoltaic Cells & Ambient, Uncontrollable & 0.01$-$100 mW/cm$^2$ & Perennial source, High output voltage & low efficiency during cloudy days, Unavailable in night & Healthcare, Industry, Military, Transportation & \bluecite{magno2016autonomous, wu2017autonomous, satharasinghe2020investigation} \\ \hline

\textit{Artificial Light} & Photovoltaic, Electroluminescence & Ambient, Partially controllable & 10$-$100 $\mu$W/cm$^2$ & Abundant in indoor, High output voltage & Low conversion rate, low power density & Healthcare, Underground infrastructure, Pole areas & \bluecite{mishu2020prospective, hsiao2020pilot, sharma2020indoor} \\ \hline

\textit{Radio Frequency} & Rectennas & Particularly predictable, Partly controllable & 1 $\mu$W/cm$^2$ & Allows mobility, Abundant in urban lands & Low power density & Healthcare, Industry, Energy, Wireless base stations & \bluecite{mishu2020prospective, vital2019textile, wagih2020wearable} \\ \hline

\textit{Acoustic Waves} & Triboelectric, Piezoelectric, EMI & Non-ambient, Controllable, Predictable & 3-100 $\mu$W/cm$^2$ & Ubiquitous influence, Adaptable to capture different frequencies & Low power density, Frequency dependency & Healthcare, Industry, Roadways & \bluecite{fan2015ultrathin, yuan2018low, ma2021metamaterial, ma2020acoustic} \\ \hline

\textit{Water} & Capacitive & Ambient, Partially Predictable, Partially Controllable & 50 $\mu$W/cm$^2$ & Continuous operation, Abundant in nature & Depends on water pressure, Position specific, Unavailable in closed area & Underwater IoNTs, Water-based vehicles & \bluecite{kwon2014effective, zhao2022soft} \\ \hline

\textit{Wind} & Piezoelectric, Electrostatic, Triboelectric, Electromagnetic & Ambient, Predictable, Uncontrollable & 0.02$-$1.5 mW/cm$^2$ & Always Available & Fluctuating density and speed & Healthcare, Industry, Military, & \bluecite{fu2020overview, zhang2019theoretical} \\ \hline  

\textit{Body Motion} & Triboelectric, Piezoelectric, Electromagnetic, Electrostatic & Controllable, Partially Predictable & 1.5$-$125 $\mu$W/cm$^2$ & Multiple working modes, Light weight & Output power varies according to motion & Healthcare, Industry, Transportation & \bluecite{mishu2020prospective, KIM2020104904, SUN2020105035}\\ \hline

\textit{Body Heat} & Thremoelectric, Pyroelectric & Ambient (human body), Controllable, Partially Predictable & 7$-$220 $\mu$W/cm$^2$ & Wearable technology, Simple structure & Low power density, Dependence on temperature gradient & Health, Wearable technology, Industry, Military, Transportation & \bluecite{ivanov2020design, suarez2017flexible}\\ \hline

\textit{Body Sweat} & Electrochemical reactions & Partially Controllable and predictable & 0.1$-$1.2 mW/cm$^2$ & Highly biocompatible & Variable in sweat production rate, Limited availability & Healthcare & \cite{bandodkar2017soft, jia2014wearable} \\ \hline    
  \end{tabular}
}
\vspace{-2.1mm}
\end{table*}
\end{landscape}

\subsection{Influence of body-centric energy sources based IoHNT on IoNT communication}
Body-centric energy sources harness energy from physiological processes like body movement, heat, and biochemical reactions to power IoNT devices, which are essential for wearable and implantable applications.
IoHNTs can capture mechanical energy from body movements using piezoelectric and triboelectric nanogenerators. 
For instance, IoHNTs embedded in clothing can convert motion into power, supporting communication within a body-area network (BAN) where devices relay physiological or biochemical data.
Body heat-based IoHNTs exploit the temperature difference between the body and the environment or temperature fluctuations to generate power. 
This energy can sustain IoNTs communications, ensuring continuous operation of health monitors that transmit data via low-power protocols like BLE (elaborated in Sec. \ref{par:Communication Protocols}).
Biofuel cells convert biochemical energy from substances like glucose into electricity, powering implantable IoNT devices such as glucose monitors. 
These cells support real-time data transmission within bio-nanonetworks using energy-efficient communication protocols.

Body-centric energy sources enable robust communication within BANs.
IoNT devices can form mesh networks, dynamically routing data to ensure reliability even in the face of movement or signal disruption.
Advances in body-centric energy harvesting will enable more autonomous and reliable IoNT devices that are integral to healthcare and personal monitoring. 
These developments will facilitate continuous, real-time health data monitoring, enhancing personalized medicine and early diagnosis.

\section{Hybrid Energy Harvesting}
\label{sec:Hybrid Energy Harvesting}
Numerous energy sources exist to generate electricity. 
However, the sources are not always available or convenient to extract electrical power. 
For example, if the energy harvesting technique depends only on the heat generated from the activities of a human when the action is ceased, the existence of the energy source disappears as well. 
Hence, it depends on the continuity of the activity, which cannot last forever. 
For that purpose, hybrid EH methods are proposed, which carry huge potential not only because of the increase in the duration of energy generation but also due to the possibility of different methods complementing the disadvantages of other methods. 
Moreover, if combined successfully, supplying energy from various sources will help the system's efficiency and functionality to increase \text{{\color{blue} \cite{7111275, abcd}}}. 
Some of the hybrid systems and their output power performances are given in Table \ref{tab:hybr}.

\begin{table}[t!]
\centering
    \caption{Performances of Hybrid EH Systems.}
    \label{tab:hybr}
        \begin{tabular}{|>{\centering\arraybackslash}m{2.3cm} |>{\centering\arraybackslash}m{2.2cm} |>{\centering\arraybackslash}m{2.2cm} |>{\centering\arraybackslash}p{0.8cm}|}
        \hline
        \textbf{Hybrid Source}  & \textbf{Materials} & \textbf{Output Performance}  & \textbf{Ref.} \\ \hline
        Piezoelectric-Electromagnetic & NdFeB Magnet & 332 $\mu$W & \text{{\color{blue} \cite{challa2009coupled}}} \\ 
        \hline
        Piezoelectric-Electromagnetic & PZT-NdFeB Magnet & 1.4 $m$W & \text{{\color{blue} \cite{xu2016novel}}} \\ 
        \hline
        Piezoelectric-Electromagnetic & PZT-NdFeB  Magnet &  52.86 $\mu$W & \text{{\color{blue} \cite{khan2016hybrid}}} \\ 
        \hline
        Piezoelectric-Triboelectric & (PVDF-TrFE)-PDMS &  75.11 $\mu$W/cm$^{2}$  & \text{{\color{blue} \cite{wang2017transparent}}} \\ 
        \hline
        Piezoelectric-Triboelectric & P(VDF-TrFE)-PTFE & 10.88 mW \par 6.04 mW/cm$^{2}$  & \text{{\color{blue} \cite{zhao2019hybrid}}} \\ 
        \hline
        \end{tabular}
\vspace{-1.5mm}
\end{table}  

As there are various sources to harvest energy, the number of combinations of hybrid EH systems seems limitless initially. 
However, limitations exist for those hybridizations. 
Hence, in this review, we will only discuss the practical combinations. 
The hybrid IoHNTs can be divided into single-source and multi-source harvesting. 
In single-source, although more than one energy source is used for EH, the source type does not change, whereas, in multi-source EH, the number of types of energy sources is more than one.

\subsection{Single-Source Hybrid EH}
In single-source hybrid EH, although various energy harvesting techniques exist together, they all depend on a single energy source. 
As one type of input source can be utilized in different EH principles, exploiting all will be more efficient instead of focusing on just one principle. 
For instance, heat is produced as a byproduct in processes such as sunlight exposure, electrical generation in a circuit, and bodily activity. 
Hence, it is reasonable to build hybrid systems exploiting solar energy and heat simultaneously for better efficiency, as described in \text{{\color{blue} \cite{zhang2013integrated, tan2010energy}}}.
Therefore, even though one input exists by integrating different EH principles, it is possible to increase the total electrical power generated.  
The most common type of single-source hybridization is the kinetic energy sources \text{{\color{blue} \cite{https://doi.org/10.1002/adma.201707271}}}.
Generally, the combinations are binary. 
Some examples of these kinetic binary combinations are piezoelectric–triboelectric \text{{\color{blue} \cite{chen2017flexible, suo2016piezoelectric, li20143d}}}, piezoelectric–electrostatic \text{{\color{blue} \cite{eun2014flexible}}}, electromagnetic–triboelectric \text{{\color{blue} \cite{wang2016fully, gupta2017broadband}}}, and piezoelectric-electromagnetic \text{{\color{blue} \cite{shan2013new, toyabur2018multimodal, yang2010hybrid}}}.

However, although rare, due to the increasing complexity, trilateral combinations exist as well, such as the combination of piezoelectric, electromagnetic, and triboelectric EH technique, elaborated in \text{{\color{blue} \cite{singh2021synchronous, he2017low}}}.
A single-source hybrid IoHNT can simplify power management since it deals with energy fluctuations from a single type of ambient source, which is beneficial over a multi-source hybrid system.
Using multiple techniques to extract energy from a single source (such as mechanical movement or body heat), these energy harvesters can capture a more significant amount of energy than using only one approach. 
This results in enhanced efficiency and continuous power generation in IoHNT, which is required to wirelessly energize IoNTs.
Developing an EH that utilizes many techniques to gather energy from a single source might result in smaller and more integrated solutions, which is essential to make IoHNTs more power-dense.

\subsection{Multi-Source Hybrid EH}
Generally, different energy sources exist in the same environment simultaneously, such as heat and kinetic energy, or they replace each other places accordingly. 
For example, when the sun shines, it is possible to obtain electrical power from sunlight; however, when the rain starts, sufficient daylight no longer exists. 
With the help of multi-source hybrid EH, it is possible to cultivate energy from the sunlight and raindrops \text{{\color{blue} \cite{YOO2019424, JEON2015636}}}.  
Sunlight and kinetic energy can also be a hybrid source for EH, as indicated in \text{{\color{blue} \cite{gambier2011piezoelectric}}}. 
Besides raindrops, the motion of the waves can be hybridized with sunlight for EH \text{{\color{blue} \cite{zhong2016graphene}}}. 
Another type of hybridization combines solar energy with acoustic, in which the piezoelectric nanogenerators and photovoltaic cells are combined \text{{\color{blue} \cite{xu2011compact}}}. 
Although sunlight is more common in hybrid systems due to its higher power density and availability with respect to artificial lights, it is also possible to use indoor light to scavenge energy, as described in \text{{\color{blue} \cite{tan2010energy}}}, where both the energy of indoor light and the vibrations are used. 
As mechanical energy can be harvested in two different ways, triboelectric and piezoelectric effects, this approach has numerous applications as well. 
To illustrate, thermoelectric and triboelectric nanogenerators are used for the hybrid structure in \text{{\color{blue} \cite{kim2016triboelectric}}}, whereas instead of the triboelectric effect, the piezoelectric effect is exploited in \text{{\color{blue} \cite{yang2017hybrid}}}. 
On top of these examples, in literature, piezoelectricity and magnetic field combinations exist for EH \text{{\color{blue} \cite{xu2018hybrid}}}.

Until now, multi-source EH techniques have used different materials to harvest energy sources. 
However, it is also possible to harvest energy with one material type, provided that this material exhibits the features of more than one energy conversion mechanism. 
For example, as all pyroelectric materials are also piezoelectric, it is possible to hybridize these conversion principles by using only one type of material \text{{\color{blue} \cite{bowen2014pyroelectric, kang2016thermal}}}. 
Lastly, biofuel cells and triboelectric nanogenerators are used concurrently to harvest the biomechanical and biochemical energy together in \text{{\color{blue} \cite{li2020hybrid}}}. 
Similarly, the temperature gradient and biochemical energy are utilized to build a hybrid EH system in \text{{\color{blue} \cite{7940053}}}. 

The examples for hybrid structures are not limited to the above examples. Various combinations and improved versions of the mentioned examples exist. 
Moreover, ongoing research on individual methods will contribute to the improvement of hybrid systems. 
Hybrid IoHNTs are capable of harnessing energy from several sources, including light, heat, vibrations, and electromagnetic fields. 
This is especially advantageous for IoNTs located in close proximity to the human/plant/vehicle body since they may come into contact with several types of energy, such as body heat, movement, and surrounding light.
Hybrid IoHNTs may utilize energy from several sources to ensure a steady and dependable power supply for IoNTs. This allows IoNTs to operate successfully even in situations when one energy source is lacking or not accessible.
It provides the ability to adjust to such alterations, guaranteeing a consistent energy provision in diverse situations, such as fluctuating levels of operation, modifications in temperature, and shifting light conditions.

\subsection{Influence of hybrid energy harvesting on IoNT communication}
Hybrid IoHNTs can be categorized into single-source and multi-source strategies, each influencing communication, network topology, and quality of service (QoS) for IoNTs.
Single-source IoHNTs involve using multiple instances of the same energy type, like several solar panels or piezoelectric generators. 
This uniform approach simplifies energy management and supports consistent network topology and meshing, where devices share similar energy profiles. 
It facilitates the implementation of energy-efficient IoNT communication protocols, ensuring stable QoS with predictable energy availability.
Multi-source harvesting leverages different types of energy, such as combining solar, kinetic, and thermal sources. 
It enhances network resilience and adaptability, allowing devices to maintain communication links even when one energy source is insufficient. 
Multi-source IoHNTs support dynamic meshing and adaptive network topologies, which are crucial for maintaining high QoS in heterogeneous environments.

As IoNT networks evolve, integrating single-source and multi-source energy harvesting will enhance the resilience and efficiency of IoNT networks, enabling more autonomous and reliable operation in diverse conditions. 
Additionally, advances in AI-driven energy management and developing 6G communication protocols will further optimize energy usage, allowing IoNT devices to function seamlessly in increasingly complex and demanding applications.

\section{Wireless Power Transmission from IoHNT to IoNT}
\label{sec:Discussion}
Due to size constraints, IoNTs are deprived of a significant power source; hence, their longevity is compromised. 
Moreover, using IoHNTs directly on the IoNTs often becomes challenging due to their size limitation and placement location.
Hence, an efficient  Wireless Power Transfer (WPT) is required to transfer the harvested power to IoNTs.
This section outlines WPT mechanisms that can facilitate wirelessly transmitting harvested energy to IoNTs.
Some widely used WPTs are RF-Based Wireless Power Transfer (RFWPT), Optical Wireless Power Transfer (OWPT), Ultrasonic Wireless Power Transfer (UWPT), and Near-field Coupling (NFC).
\subsection{RF-Based Wireless Power Transfer (RFWPT)}
RFWPT uses radio frequency waves to transmit energy wirelessly across varying distances from a few centimeters to hundreds of meters.
It converts the electrical energy into RF electromagnetic (EM) waves and transmits via a transmitting antenna; that EM wave is captured by the receiving antenna integrated with the desired IoNTs.
The efficiency and range of RFWPT are influenced by the frequency of the RF waves, power output, antenna design, and environmental conditions.  
RFWPT is particularly useful to power IoNTs as it can efficiently handle the low-power transmission \bluecite{luo2019rf}.
Moreover, due to the omnidirectionality of the power transferring feature, it can be quickly adopted in different situations for powering multiple IoNTs at a time in industry, medical, military, health, and even for in-home use.
\subsection{Optical Wireless Power Transfer (OWPT)}
OWPT uses a method to transmit energy using optical beams, typically laser beams.
It uses highly focused laser beams to carry energy from the IoHNT through an open region to a photovoltaic receiver placed on IoNT to convert the light energy into electrical energy. 
This technique is advantageous where a direct line of sight (LOS) can be maintained, as it offers minimal dispersion and high energy transfer efficiency due to the precision and directionality of laser beams \bluecite{fakidis2016indoor}.
Due to safety concerns as laser rays can badly affect biological cells, OWPT is recommended for powering IoNTs in closed autonomous industries, space-based IoNTs, and Robotics.
In order to power multiple IoNTs at a time, multiple OWPT modules need to be integrated with IoHNT.
\subsection{Ultrasonic Wireless Power Transfer (UWPT)}
UWPT is a promising technology that uses ultrasonic waves to transmit power wirelessly over a short distance.
The use of UWPT is suitable in an industry where IoNTs are located inside a metallic barrier or behind a metallic structure; the IoHNT harvested power can be easily transmitted to those devices without any physical penetration \bluecite{kar2020performance}.
It can be used to power medical implants and wearable IoNTs, along with the adaptation in industrial and transportation sectors.
\subsection{Near-field Coupling (NFC)}
NFC operates over relatively short distances (up to a few meters). 
It can be broadly classified into three: Magnetic Resonance Coupling (MRC), Inductive Coupling (IC), and Capacitive Coupling (CC).
\subsubsection{Magnetic Resonance Coupling (MRC) and Inductive Coupling (IC)}
MRC can be considered as an enhanced Inductive Coupling technique (IC).
The basic operating principles of IC and MRC are the same.
They utilize magnetic fields to transmit energy between two resonant circuits wirelessly. 
It is mostly used for short-range (typically up to one meter) power transmission techniques and sometimes for medium-range transmission (up to several meters). 
It comprises two components: one transmitter coil attached with IoHNT and one receiver coil attached with IoNT. MRC requires proper tuning to transmit power, but IC does not \bluecite{ali2021comprehensive}, which makes MRC for use in sophisticated applications.
This technique can be well suited for transmitting power over a broader area, like an MRC attached to an IoHNT placed on a body can transmit power to multiple IoNTs placed inside the body.
Similarly, it can be used for environmental monitoring and plant monitoring.
\subsubsection{Capacitive Coupling (CC)}
CC utilizes electric fields between two parallel conductive plates to transfer energy.
When an alternating voltage is created at the transmitter plate, it makes a time-varying electric field through the dielectric and extends to the receiver electrode. 
Like MRC, CC must also be tuned to its optimal frequency and impedance conditions, depending upon the distance between the electrodes, dielectric constant, and the size of the plates \bluecite{wang2022research}.
CC can be used in industrial, consumer applications, medical, and transportation sectors \bluecite{wang2022research}.

\begin{table}[t!]
    \centering
    \caption{Comparison among WPT techniques}
    \begin{tabular}{|>{\centering\arraybackslash} m{1.5cm}  |>{\centering\arraybackslash} m{1.5cm}  |>{\centering\arraybackslash} m{1.2cm}  |>{\centering\arraybackslash} m{1.3cm} |>{\centering\arraybackslash} m{1.4cm} |>{\centering\arraybackslash} m{1.2cm}  |>{\centering\arraybackslash} m{1.7cm}  |}
    \hline
    \textbf{Technique} & \textbf{Field Type} & \textbf{Range} & \textbf{Efficiency} & \textbf{Operating Frequency} & \textbf{Multi Casting} & \textbf{IoNT powering areas} \\ \hline
    \textit{RFWPT} & EM & Medium/ long & Low & RF & Present & Industry, Medical, Military, In-home \\ \hline
    \textit{OWPT} & EM & Medium/ long & Medium & $\geq$THz & Absent & Industries, Space\\ \hline
    \textit{UWPT} & Ultrasound & Short & Medium & MF & Present & Medical, Environment\\ \hline
    \textit{MRC} & Magnetic & Short & High & HF & Present & Medical, Environment, Industry\\ \hline
    \textit{CC} & Magnetic & Very short & High & LF \& HF & Absent & Industry, Transportation\\ \hline
    \end{tabular}
    \label{tab:WPT}
\end{table}

\setcounter{paragraph}{0}
These well-known WPT techniques can be used as a medium between IoHNTs and IoNTs. 
Incorporation of this medium will lead to prolonged operation of IoNTs without any maintenance.
A detailed comparison among all the discussed WPT techniques is made in Table \ref{tab:WPT}.
However, depending on the discussion, RFWPT is more favorable where multiple IoNTs are in a given range. Still, significant research is required to achieve favorable power transmission efficiency. 
For a limited number of IoNTs (1-2) in a short range, NFC is suitable as a WPT technique.
Hence, RFWPT and NFC need enhanced attention as suitable WPT techniques for powering IoNTs and an open area of research to build a compact IoHNT.
The mechanism of WPT widely depends on the functional duration of the IoNTs, classified into two:

\textit{(i) Continuous Wireless Power Transfer (CWPT)}: If an IoNT is operating continuously, that requires a continuous power supply from IoHNT. There is no need for a power storage system on board the IoNT as power is continuously being transmitted by IoHNT.

\textit{(ii) On-Demand Wireless Power Transfer (ODWPT)}: Depending upon the condition or once the threshold condition is triggered, it can ask for the power supply through a communication link. This kind of IoNT requires a tiny energy-storing device on board for the threshold detection and set-up of a communication link.

Furthermore, ODWPT can be classified into two depending upon the IoHNT architecture: (i) \textit{Harvest-Transfer}: Depending on the activation of the communication link between IoNT and IoHNT, energy is harvested and instantly transferred and (ii) \textit{Harvest-Store-Transfer}: In this technique, IoHNT is associated with a storing device like a supercapacitor, and transmit power to IoNT depending on the requirement of IoNT.
For example, an IoNT, with a tiny energy-storing device, is designed to be functional only when a threshold of a system is reached, like an IoNT designed to be functional and transfer the ECG and pulse data to medical personnel through the gateway only when it reaches a certain threshold, there is a need of having a communication link between IoHNT and IoNT seeking for power to its nodes. The IoNT's storage device will only be used to set up the communication link between IoNT and IoHNT. Once the communication link is set up, IoHNT can wirelessly energize all the nodes of IoNT. 

\begin{table}[t!]
    \centering
    \caption{Comparison among IEEE 802.15.1/3/4 Protocols}
    \begin{tabular}{|>{\centering\arraybackslash} m{1.7cm}  |>{\centering\arraybackslash} m{2.2cm}  |>{\centering\arraybackslash} m{2.2cm}  |>{\centering\arraybackslash} m{2.2cm}  |}
    \hline
    \textbf{Feature} & \textbf{IEEE 802.15.1} & \textbf{IEEE 802.15.3} & \textbf{IEEE 802.15.4}  \\ \hline
    Technology & Designed for Bluetooth  & Designed for HR-WPAN & Designed for LR-WPAN \\ \hline
    Frequency  & 2.4 GHz & 3.1-10.6 GHz  & 2.4 GHz, 915/868 MHz  \\ \hline
    Data rate & $\leq$ 1 Mbps & 110 Mbps to few Gbps  & $\leq$ 250 kbps \\ \hline
    Power (Tx/Rx) & 102.6/84.6 mW & 750/750 mW & 74/81 mW \\ \hline
    Protocol Complexity & Very high (188 Primitives) & Moderately high (106 primitives) & low (48 primitives) \\ \hline
    Range & $\leq$ 10 meters & 10-100 meters & 10-100 meters \\ \hline
    \end{tabular}
    \label{tab:IEEE}
\end{table}

\section{Impact on IoNT Communication}
\label{sec:influ}
\subsection{Communication Protocols}
\label{par:Communication Protocols}
Communication protocols that are being used for IoNTs and the performance enhancement that IoHNTs can bring in IoNTs are discussed here.
\textit{IEEE 802.15} group defines some wireless personal area network (WPAN), namely 802.15.1 \bluecite{xiao2009emerging}, 802.15.3 \bluecite{xiao2009emerging}, and 802.15.4 \bluecite{ergen2004zigbee, xiao2009emerging}; typically classified in Bluetooth, high rate WPAN (HR-WPAN), and low rate WPAN (LR-WPAN). 
A comparison of these communication protocol standards is given in Table \ref{tab:IEEE} \bluecite{lee2007comparative}. 
Furthermore, Bluetooth Low Energy (BLE) is a non-IEEE communication protocol for short-range low energy power communication. 
It remains in sleep mode and activates only when a connection is initiated for data transfer, significantly extending the device's battery life.
The data rate of older versions of BLE, Bluetooth 4.0 is up to 1 Mbps; later versions (Bluetooth 5.0 onwards) offer up to 2 Mbps \bluecite{park2020adaptable}.  

IEEE 802.15.4 is typically used for the IoNTs due to its low power consumption, low data rate transmission, and low quality of service. 
The incorporation of IoHNTs leads to the availability of continuous power at the nodes of IoNTs.
Available power can enhance communication data rates by providing frequent updates and high-quality data transmission.
With IoHNTs, IoNTs can afford a higher duty cycle, which allows them to be active more often. This leads to faster response time and higher throughput.
Two more types of communication protocols are available for IoNTs: Long Range Wide Area Networks (LoRaWAN) and IPv6 over Low-Power Wireless Personal Area Networks (6LoWPAN). LoRaWAN is an open-source protocol for long-range communication at a very low data rate ($\leq$50 kbps) \bluecite{de2017lorawan}. 6LoWPAN is a protocol that transmits IPv6 packets over IEEE 802.15.4 protocol with a data rate similar to LoRaWAN \bluecite{huang2015implementation}.   
Hence, these protocols can not be used in those IoNTs where high-rate data transmission is a priority. 
For that, IEEE 802.15.1/3 needs to be favored over IEEE 802.15.4. 
The incorporation of IEEE 802.15.1/3 protocols leads to a requirement for high power, which can be supplied by IoHNTs.
Incorporating IoHNT will lead to continuous operation of IoNTs, and real-time data can be transmitted to desired places with high data rate, high precision, and high quality at a very high speed, which will positively impact all sectors, like health, agriculture, industrial, military, and so on.

\subsection{Network Topology}
\label{mesh}
In IoNTs, network topology is crucial for maintaining connectivity among tiny devices. 
Three main types of topologies are widely explored in the context of nanodevices$-$ (i) \textit{Star Topology}: In this communication topology, all the nodes transmit data to a central gateway, and that central gateway is responsible for data processing and routing \bluecite{Jiang2015/11}. 
This is an energy-efficient technique; however, interruption in any communication channel can significantly impact the output data. 
(ii) \textit{Tree Topology}: Tree topology can be considered as an introduction to hierarchy in star topology; it is beneficial where there is a need for data analysis at intermediary stages \bluecite{Jiang2015/11}. 
However, similar to a star topology, it suffers reliable output for any impact in any communication channel. 
(iii) \textit{Mesh Topology}: Mesh topology is the most interconnected topology where most of the nodes are connected with each other (if all the nodes are connected with each other, that is called fully connected topology).
Mesh topology is widely used where communication and data reliability are highly preferred. 
However, a detailed mesh topology analysis is out of this paper's scope.
A comprehensive analysis of the mesh network has been done in \bluecite{1509968}.
This ensures data reliability even with interruptions with a single communication channel by routing the information through other nodes. 
However, Mesh topology requires high power as each node is connected to more than one node, and this mesh structure also makes the whole system computationally extensive. 
The continuous power supply from IoHNTs through WPT can significantly boost the operation of IoNTs by pushing toward the Mesh topology and providing a highly reliable, uninterrupted communication architecture. 

\subsection{Data Transmission Strategies}
In IoNTs, depending on the purpose and importance, the data transmission strategies are adopted as follows:
(i) \textit{Scheduled data transmission}: In this type, nodes can transmit data only at a predefined specific time. It needs low power and is used in environmental monitoring and health monitoring, such as blood pressure monitoring at night IoNTs. 
(ii) \textit{Condition-based data transmission}: In this data transmission type, IoNT nodes can only transmit data after meeting a specific condition or a threshold is reached. These types of IoNTs require energy that depends on the estimated time of operation, from low to moderately high power. It can be used in waste management, infrastructure, healthcare, etc.
(iii) \textit{Continuous data transmission}: This type of data transmission requires high power due to its continuous operation.
IoNTs with continuous data transmission are crucial where real-time data analysis and immediate response are required. 
For example, in chemical manufacturing or nuclear power plants, IoNTs may continuously transmit data about temperature, pressure, and radiation levels to officials through a gateway to ensure that they remain within safe limits and prompt action can be taken to prevent any probable accident. 
It can also be incorporated into infrastructure, disaster management, transportation, environment, and health monitoring.

\subsection{Quality of Service (QoS)}
Incorporation of IoHNTs in IoNTs offers a significant improvement in QoS discussed in the following $-$
(i) \textit{Higher data transmission rate}:
Continuous power transmission enables IoNTs to perform more complex tasks and transmit higher data rates. 
Due to the continuous power supply, the IEEE 802.15.4 communication protocol's data rate can be significantly enhanced.
Incorporation of 802.15.3 leads to a high data rate transmission, as discussed earlier, which can significantly promote QoS. 
This can lead to bandwidth-intensive applications, e.g., video streaming in medical procedures or high-resolution sensor data monitoring in industrial processes. 
(ii) \textit{Reduced latency and Reliability}:
The availability of continuous power with IoNTs can lead to the use of sophisticated algorithms for data processing and transmission, as well as error detection and enhancement in communication protocols. This ensures a continuous, reliable, and real-time data transmission with negligible latency. It is crucial in healthcare, industrial, and accident prevention.
(iii) \textit{Traffic management}:
The continuous power supply from IoHNTs leads to the incorporation of algorithms like priority queuing \bluecite{Walraevens2011}, a traffic shaping technique. Incorporation of this algorithm in healthcare IoNTs can lead to sending data with critical alerts on a priority basis over routine data updates, which can help doctors diagnose and treat quickly.
(iv) \textit{Security and Privacy}:
Security and privacy is a major concern in communication. However, the IEEE 802.15.4 protocol with ZigBee and IEEE 802.15.3 are considered to be secured. Still, the incorporation of IoHNTs leads to the addition of complex encryption methods and continuous security updates in IoNTs.

Overall, the incorporation of IoHNTs to power IoNT will significantly shift the concept of powering nanodevices, which will lead to an \textit{interconnected nano world}.

\section{Future Scope of IoHNTs}
\label{sec:future}
\begin{figure*}[!t]
\centering
{\includegraphics[width=0.95\textwidth]{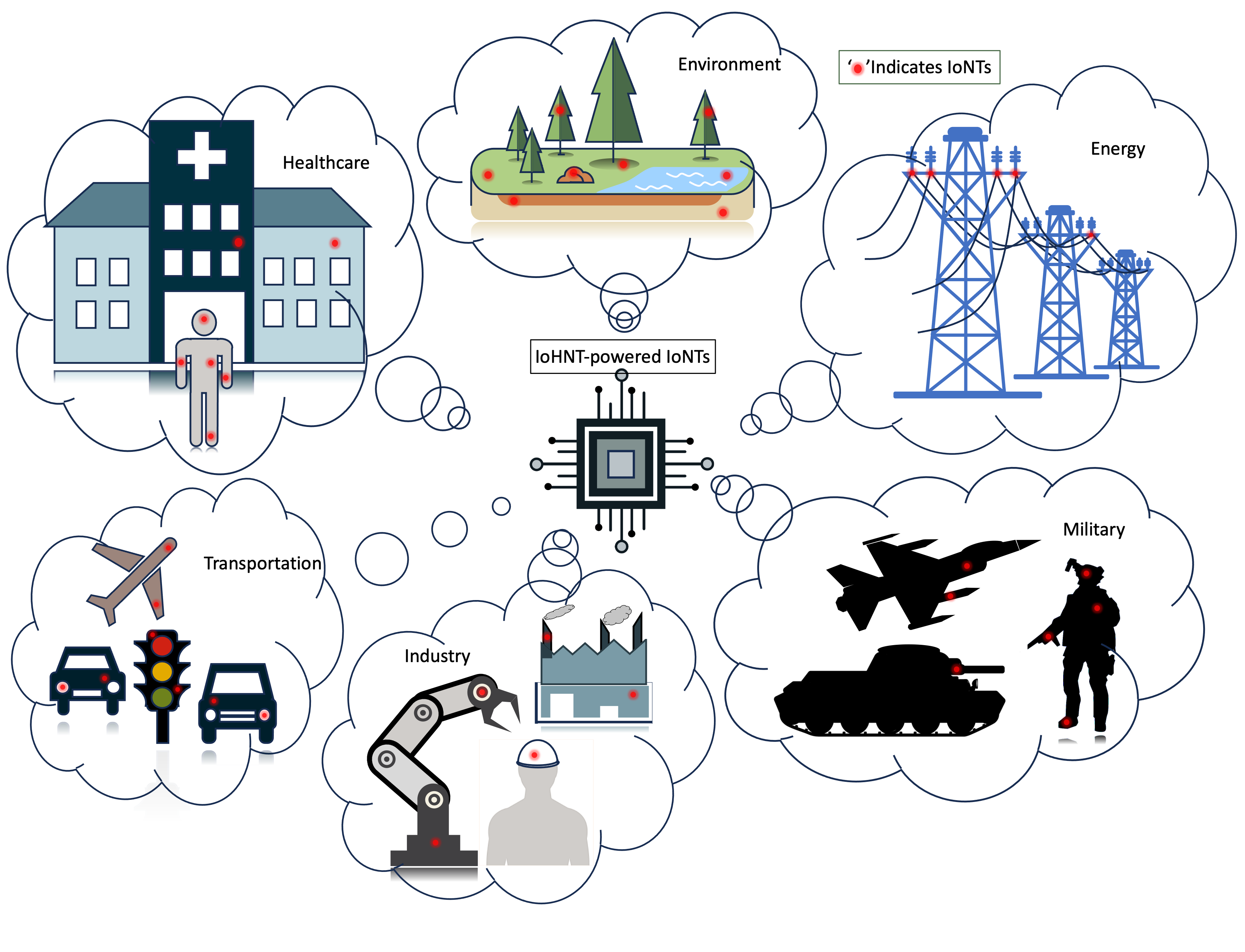}}
\caption{Future of IoHNT-powered IoNTs.}
\label{future}
\end{figure*}

Introducing IoHNT-based IoNTs represents a significant shift in sustainable communications in nanodevices. IoHNTs mitigate the power limitation of traditional IoNTs and facilitate the development of autonomous and spontaneous communication systems. Here, we explore the future scope of IoHNTs by detailing prospective advancements and applications across diverse sectors. A futuristic overview of IoHNT-based IoNTs in illustrated in Fig. \ref{future}.

\subsection{Environmental Monitoring}
An important application of IoHNTs is the development of autonomous environmental monitoring systems. The harnessed energy of IoHNTs is wirelessly transmitted to IoNT sensor nodes placed over large and often inaccessible locations. For example, IoNTs placed in soil in a forest could monitor soil moisture and environmental toxins to detect early signs of drought or pollution. Through the continuous power delivery by IoHNTs, these IoNTs can operate for a long time and send real-time data to operating personnel or the cloud through a gateway for further processing, which is crucial for environmental monitoring and disaster prevention.

\subsection{Healthcare Sector}
IoHNTs can significantly impact healthcare by enabling the continuous operation of IoNTs used for medical diagnostics and treatment.
IoHNTs can scavenge energy from body heat or kinetic energy from body movements and transmit it continuously to IoNTs for monitoring glucose levels, blood pressure, ECG, etc. One groundbreaking application of IoHNTs will be where full-body monitoring is required, and multiple IoNTs can be placed at different body locations to monitor various types of biomarkers or cancer progression.
Nonetheless, with the fast progress of nanotechnology, IoNTs can be modeled to capture real-time time-lapse images of an organ or a position to monitor the progress of a tumor or cancer directly. All these IoNTs require a continuous power supply to run for a long time; the power requirements can be easily tackled through WPT from IoHNTs to IoNTs.

\subsection{Energy Distribution}
IoHNT can revolutionize the way smart grids operate. For instance, IoNTs can be deployed throughout the smart grids to monitor real-time electrical parameters like voltage, current, frequency, etc. 
Incorporation of such IoNTs will lead to the detection of failure and maintenance requirements, which will reduce downtime and maintenance costs. 
Moreover, IoNTs integrated with various appliances or devices can manage the dynamic loads by continuously sending the mentioned electrical parameters to operators. 
All these IoNTs require a continuous power supply, and numerous IoNTs need to be placed in a particular area for reliable data acquisition and transmission. IoHNTs can be used to power those IoNTs, leading to green energy-run devices.
Furthermore, an IoHNT can transmit energy to nearby IoHNTs depending on the power requirement and availability. 

\subsection{Transportation Sector}
IoNTs can continuously monitor traffic congestion and do real-time data streams about traffic conditions, road occupancy, and vehicle speeds across vast networks. 
The control room can respond to the streamed data and make traffic predictions and management decisions to reduce congestion and improve flow. 
The  Multiple nodes of IoNTs can be powered through IoHNTs for uninterrupted operation without any risk of downtime due to power cuts. 
Moreover, IoHNTs can power all the nodes of IoNTs in different structures like bridges, tunnels, street-side buildings, and pavements to transmit real-time data of any structural failure or hazards. 
Furthermore, IoHNT can significantly collaborate to improve the Internet of Vehicular Things (IoVT) by providing continuous power to all the IoNT nodes on/in a vehicle, facilitating improvement in Advanced Driver-Assistance Systems (ADAS) with IoNTs for collision detection or parking detection, vehicle health monitoring, tire pressure monitoring, and anti-theft technology.

\subsection{Defence Sector}
Incorporating IoHNTs in IoNTs offers remarkable benefits in the military sector by continuously monitoring soldiers' health and surveillance. 
IoHNT-powered wearable IoNTs can monitor health and environmental conditions. 
In any battlefield, it is vital to constantly monitor these data to support the troops with immediate response. 
IoNT-enabled continuous surveillance in remote locations to detect movements and hazardous conditions is crucial to detect any activity in sensitive areas and provide an early warning without any frequent maintenance operation. 
Mesh-based IoNT architectures is suitable for this sector, where IoHNTs can power all the available IoNTs within their communication range. 
Overall, IoHNT-powered IoNT technologies significantly bolster military efficiency, safety, and strategic execution.

\subsection{Industry} 
IoHNT-based IoNT can offer real-time health and operational monitoring of industrial equipment, facilitating maintenance prediction and reducing downtime by identifying potential failures. 
In some sectors like chemical or nuclear, IoNTs can monitor environmental conditions such as toxins in the air, air quality, temperature, etc., and transmit to officials remotely, ensuring the safety of the industrial sector.
Moreover, IoNTs can track material production from receipt to delivery in the supply chain and ensure the condition and timely delivery of goods.
Furthermore, IoNTs can do nano-level inspections of goods and detect production defects, which can help produce high-quality products. 
All the IoNTs in an industry can be powered with the help of a few IoHNTs, depending upon the area of the industrial premises. 
The all-inclusive incorporation of IoHNT-based IoNTs in the industry revolutionizes productivity, safety, and efficiency, leading to smart and advanced industrial operations.

These are some examples where IoHNT-based IoNTs can be easily adopted but are not strictly limited to these. It has a high impact across all sectors, from molecules to the universe.

\section{Conclusion}
\label{sec:Conclusion}
The emergence of the IoNT signifies a substantial advancement in the field of intelligent technology and the IoT. 
The integration of nanotechnology with the connection of IoT, known as IoNT, offers revolutionary applications in almost all sectors, including healthcare, environmental monitoring, transportation, industrial, etc. 
Nonetheless, the effectiveness and long-term viability of IoNTs are greatly influenced by the progress made in the EH module, IoHNT. 
This is because these tiny devices need a consistent and dependable power supply in order to operate efficiently. 
One of the primary challenges IoNT devices face is their limited energy capacity. 
IoHNT provides a practical answer by transforming energy from different sources in the environment into electrical energy and transmitting it wirelessly to IoNTs, which can be used to power nanodevices. 

Considering the increment in urbanization, predicted to be 70\% by 2050, the Internet of Everything (IoE) will become the most important feature for continuous and smooth operation, and IoE will be backed by the massive implementation of IoNTs \bluecite{miraz2018internet}. 
Hence, having compact and efficient IoHNT is crucial to power those IoNTs simultaneously and continuously. 
This paper presents a compact literature on EH and WPT techniques, the two main sections of which IoHNT is made up. Environmental sources and in-body sources are considered to be potential sources of energy that can be harvested by the IoHNTs. 
The harvested power is then transmitted to IoNT nodes through proper WPT techniques for efficient and continuous operation, which will effectively improve the whole communication architecture of IoNTs. 
Considering the EH technologies and the WPT techniques, using the hybrid energy harvester and the RFWPT/NFC would be deemed to be the best combination for making IoHNT. 
However, till today, there is no evidence of fabricated IoNTs; it is an ongoing research domain. 
Hence, further research should be prioritized to obtain a suitable integration technique in order to maximize the power transmission from IoHNTs to IoNTs. 
It will inspire scholars to think about and work toward the $Next$ power revolution in IoNTs, providing unparalleled efficiency and simplicity.
We anticipate that this study will expand the range of EH methods and enable the development of more IoNTs with increased reliance on IoHNT.

\bibliographystyle{elsarticle-num} 
\bibliography{references}

\end{document}